\documentclass[twocolumn]{aastex61_}

\usepackage{url}
\usepackage{hyperref}

\shorttitle{Nightside Winds of Venus with Akatsuki/IR2}
\shortauthors{Peralta et al.}
\begin{document}

\title{Nightside Winds at the Lower Clouds of Venus with Akatsuki/IR2: Longitudinal, local time and decadal variations from comparison with previous measurements.}

\correspondingauthor{Javier Peralta}
\email{javier.peralta@ac.jaxa.jp}

\author[0000-0002-6823-1695]{Javier Peralta}
\affil{Institute of Space and Astronautical Science (ISAS), Japan Aerospace Exploration Agency (JAXA) \\
3-1-1, Yoshinodai, Chuo-ku, Sagamihara, Kanagawa, 252-5210, Japan}

\author{Keishiro Muto}
\affiliation{Graduate School of Frontier Sciences, The University of Tokyo, Japan}
%\collaboration{(Manual cloud tracking of IR2 images)}
%\nocollaboration

\author{Ricardo Hueso}
\affiliation{Escuela de Ingenier\'ia de Bilbao (UPV/EHU), Bilbao, Spain}
%\collaboration{(Manual cloud tracking of VIRTIS images and Contour plots)}

\author{Takeshi Horinouchi}
\affiliation{Faculty of Environmental Earth Science, Hokkaido University, Sapporo, Japan}
%\collaboration{(Full-automatic cloud tracking of IR2 images)}

\author{Agust\'in S\'anchez-Lavega}
\affiliation{Escuela de Ingenier\'ia de Bilbao (UPV/EHU), Bilbao, Spain}
%\collaboration{(Manual cloud tracking of VIRTIS images)}

\author{Shin-ya Murakami}
\affiliation{Institute of Space and Astronautical Science (ISAS), Japan Aerospace Exploration Agency (JAXA) \\
3-1-1, Yoshinodai, Chuo-ku, Sagamihara, Kanagawa, 252-5210, Japan}
%\collaboration{(Full-automatic cloud tracking of IR2 images

\author{Pedro Machado}
\affiliation{Institute of Astrophysics and Space Sciences, Portugal}
%\collaboration{(TNG/NICS observations)}

\author{Eliot F. Young}
\affiliation{Southwest Research Institute, Boulder, CO 80302, USA}
%\collaboration{(IRTF/SpeX observations)}

\author{Yeon Joo Lee}
\affiliation{Graduate School of Frontier Sciences, The University of Tokyo, Japan}
%\collaboration{(IRTF/SpeX observations)}

\author{Toru Kouyama}
\affiliation{Artificial Intelligence Research Center, National Institute of Advanced Industrial Science and Technology, Tokyo, Japan}
%\collaboration{(IRTF/SpeX observations)}

\author{Hideo Sagawa}
\affiliation{Faculty of Science, Kyoto Sangyo University, Japan}
%\collaboration{(IRTF/SpeX observations)})}

\author{Kevin McGouldrick}
\affiliation{Laboratory for Atmospheric and Space Physics, University of Colorado Boulder, Boulder, CO 80303-7814, USA}
%\collaboration{(Discussion of results)}

\author{Takehiko Satoh}
\affiliation{Institute of Space and Astronautical Science (ISAS), Japan Aerospace Exploration Agency (JAXA) \\
3-1-1, Yoshinodai, Chuo-ku, Sagamihara, Kanagawa, 252-5210, Japan}
\affiliation{Department of Space and Astronautical Science, School of Physical Sciences, Sokendai, Japan}
%\collaboration{(Correction and calibration of IR2 images)}

\author{Takeshi Imamura}
\affiliation{Graduate School of Frontier Sciences, The University of Tokyo, Japan}
%\collaboration{(Discussion of results)}

\author{Sanjay S. Limaye}
\affiliation{Space Science and Engineering Center, University of Wisconsin-Madison, Madison, USA}

\author{Takao M. Sato}
\affiliation{Space Information Center, Hokkaido Information University, Hokkaido, Japan}
\affiliation{Institute of Space and Astronautical Science (ISAS), Japan Aerospace Exploration Agency (JAXA) \\
3-1-1, Yoshinodai, Chuo-ku, Sagamihara, Kanagawa, 252-5210, Japan}
%\collaboration{(IRTF/SpeX observations)}

\author{Kazunori Ogohara}
\affiliation{School of Engineering, University of Shiga Prefecture, Shiga, Japan}
%\collaboration{(Full-automatic cloud tracking of IR2 images)}

\author{Masato Nakamura}
\affiliation{Institute of Space and Astronautical Science (ISAS), Japan Aerospace Exploration Agency (JAXA) \\
3-1-1, Yoshinodai, Chuo-ku, Sagamihara, Kanagawa, 252-5210, Japan}

\author{David Luz}
\affiliation{Institute of Astrophysics and Space Sciences, Portugal}
%\collaboration{(TNG/NICS observations)}

\date{Published in \textit{Astrophysical Journal Supplement Series} the 7 of December 2018; original manuscript \href{http://iopscience.iop.org/article/10.3847/1538-4365/aae844/meta}{here}.}



\begin{abstract}
We present measurements of the wind speeds at the nightside lower clouds of Venus from observations by JAXA's mission Akatsuki during 2016, complemented with new wind measurements from ground-based observations acquired with TNG/NICS in 2012 and IRTF/SpeX in 2015 and 2017. Zonal and meridional components of the winds were measured from cloud tracking on a total of 466 Akatsuki images of Venus acquired by the camera IR2 using the 2.26-$\mathrm{\mu m}$ filter, with spatial resolutions ranging 10--80 km per pixel and covering from 2016 March 22 to October 31. More than 149,000 wind vectors were obtained with an automatic technique of template matching, and 2,947 wind vectors were inferred with the manual procedure. The meridional profiles for both components of the winds are found to be consistent with results from the Venus Express mission during 2006--2008, although stronger wind variability is found for the zonal component at equatorial latitudes where Akatsuki observations have better viewing geometry than Venus Express. The zonal winds at low latitudes also suggest a zonal variability that could be associated with solar tides or vertically propagating orographic waves. Finally, the combination of our wind measurements from TNG/NICS, IRTF/SpeX and Akatsuki images with previously published and based in data from 1978 to 2017 suggests variations of up to 30 m s$^{-1}$ in the winds at the lower clouds of the Venus nightside.
\end{abstract}




\section{Introduction}\label{sec:intro}
The atmospheric circulation of Venus from the surface up to the stratosphere is dominated by a retrograde zonal superrotation (h.a. RZS) which, according to both \textit{in situ} \citep{Counselman1980} and cloud tracking measurements \citep{Sanchez-Lavega2008}, attains the fastest wind speeds at the top of Venus's cloud layer \citep{Schubert1983,Gierasch1997,Sanchez-Lavega2017}. The cloud layer is also where most of the energy from the solar radiation is deposited \citep{Lee2015PSP}, so a detailed characterization of the atmospheric circulation at multiple levels of the clouds \citep{Sanchez-Lavega2008,Peralta2017GRL} seems essential to properly evaluate the sources and transport of angular momentum in the atmosphere. The wind speeds at the upper clouds of Venus (60--70 km above the surface) are customarily characterized with cloud tracking on both day \citep{Rossow1990,Belton1991,Peralta2007b,Limaye2007,Kouyama2013,Khatuntsev2013,Hueso2015,Horinouchi2018} and night sides \citep{Peralta2017NatAstro}, while the  winds at the deeper middle and lower clouds (about 48--60 km) have been evaluated following cloud features visible on the dayside of the planet on the albedo at near infrared wavelengths \citep{Belton1991,Hueso2015,Khatuntsev2017} and in the lower clouds on the nightside \citep{Sanchez-Lavega2008,Hueso2012,Horinouchi2017NatGeo} thanks to inhomogeneities in the opacity of the lower clouds which can be observed at the infrared windows at 1.7 and 2.2-2.3 $\mathrm{\mu m}$ \citep{Peralta2017Icarus}.\\
\\
After the discovery of these infrared spectral windows by \citet{Allen1984}, and before the arrival of Akatsuki \citep {Nakamura2016}, ESA's Venus Express mission (h.a. VEx)  characterized the RZS at the deeper clouds with unprecedented detail during 2006--2008. However, constraints due to VEx's polar orbit \citep{Titov2006} and the long exposure times required by the imaging spectrometer VIRTIS-M \citep{Piccioni2007ESA} limited observations to the southern hemisphere of Venus with lower quality results at equatorial and lower latitudes \citep{Sanchez-Lavega2008,Hueso2012,McGouldrick2012,Peralta2017GRL}. As a result, the circulation at the level of the deeper clouds is yet poorly characterized in the northern hemisphere where only sparse measurements exist from ground-based observations \citep{Crisp1989,Crisp1991b,Chanover1998,Limaye2006} and images from the Near-Infrared Mapping Spectrometer (NIMS) instrument during the flyby of NASA's Galileo in 1990 \citep{Carlson1991}. Since its orbit insertion in 2015 December, JAXA's Akatsuki orbiter has permitted an invaluable opportunity to study this deeper atmospheric circulation on both hemispheres thanks to its equatorial orbit \citep{Nakamura2016} and the images provided by the 2-$\mathrm{\mu m}$ camera IR2 \citep{Satoh2016}.\\
\\
In this work we present the first global measurements of the wind speeds at the nocturnal lower clouds of Venus during the first year of Akatsuki observations. A description of the observations performed by the IR2 camera, the processing of the images, navigation corrections and methods of cloud tracking are introduced in section \ref{sec:methods}. The lower clouds' morphologies, the meridional profiles of wind speeds and the possible relation between morphologies and speeds are presented in section \ref{sec:meridprofiles}. The wind speeds' dependence on the size and opacity of the clouds, the local time and an exploration of possible relations with the surface topography is studied in section \ref{sec:WindDependences}, while the temporal variation of the winds of the lower clouds is presented in section \ref{sec:TimeEvol}. Finally, the main conclusions of this work are presented in section \ref{sec:conclus}.\\

\section{Methods}\label{sec:methods}
After failing its originally-planned orbit insertion in 2010 December 7, Akatsuki orbited the Sun for 5 years and was inserted with success into a Venus orbit in 2015 December \citep{Nakamura2016}. At present, the orbiter performs a westward equatorial orbit with an apoapsis of $\sim$360,000 km, periapsis ranging 1,000--8,000 km, and a rotation period of about 10 days. Among the payload, the camera IR2 was designed to sense the deep clouds of Venus and infer information about the atmospheric compounds below the cloud layer thanks to its narrow-band filters centered at 1.74, 2.26 and 2.32 $\mathrm{\mu m}$ \citep{Satoh2016}. The scientific objectives of these filters include the study of the morphology of the clouds and their motions, the aerosols' properties or the abundance of the CO below the clouds \citep{Satoh2017}. Because of the highly eccentric orbit of Akatsuki, the spatial resolution of Venus in the IR2 images varies from 74--12 km per pixel for off-pericenter observations, to 1.6--0.2 km during pericentric ones \citep{Nakamura2016}. Unfortunately, the acquisition of IR2 images was indefinitely interrupted in 2016 December 9 when the electronic device controlling the IR1 and IR2 cameras started to experience an unstable power consumption that has persisted through the present time \citep{Iwagami2018}.\\
\\
In addition to the IR2 images, new wind measurements have been obtained in this work with ground-based images of Venus acquired after the failure of the infrared channel of the VIRTIS-M \citep{Hueso2012} instrument on VEx and before and after the time period covered by Akatsuki IR2 observations. The Near Infrared Camera Spectrometer (NICS) \citep{Baffa2001} at the Italian National Telescope Galileo (TNG) at La Palma (Canary Islands, Spain) was used to acquire images of the nightside of Venus during 2012 July 11--13 \citep{Machado2016DPS}, while the Medium-Resolution 0.8--5.5 Micron Spectrograph and Imager (SpeX) \citep{Rayner2003} at the 3-m National Aeronautics and Space Administration Infrared Telescope Facility (IRTF) was used to provide images of Venus with the K-continuum filter in 2015 September and 2017 January-February \citep{Lee2017JpGU}. The wind measurements from these ground-based images will be presented in section \ref{sec:TimeEvol}.\\

\subsection{Image processing}\label{ssec:observ-processing}
\begin{figure*}
\centerline{\includegraphics[width=40pc]{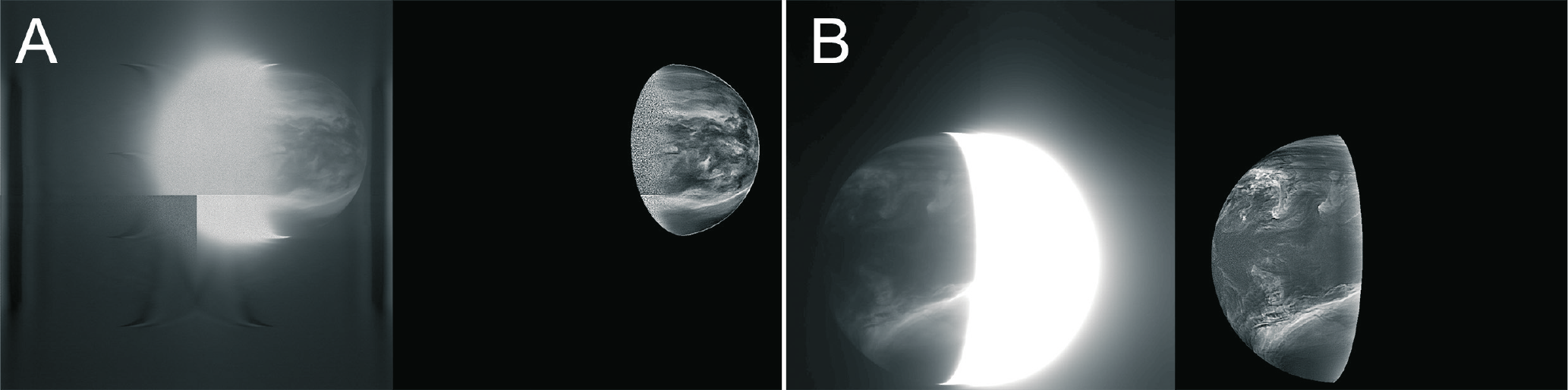}}
\caption{Examples of IR2/2.26-$\mathrm{\mu m}$ images before and after the processing procedure described in this work: (\textbf{A}) IR2 image acquired in March 25, 11:33 UT, and (\textbf{B}) IR2 image acquired in August 25, 04:03 UT. Both images have been oriented so that the north-south-west-east of Venus appears approximately oriented up-down-left-right in the image.}
\label{figure:process-images}
\end{figure*}

IR2 acquired a total of 1,671 images\footnote{Available at: \url{http://darts.isas.jaxa.jp/planet/project/akatsuki/}} of the nightside of Venus with the filters 1.74, 2.26 and 2.32 $\mathrm{\mu m}$ \citep{Satoh2017} among which $\sim$1,370 were regarded as suitable for cloud tracking. \citet[fig.~8a therein]{Satoh2017} reported a problem of light contamination in the IR2 images, consisting of the presence of \textit{halation} rings and a cross pattern extending both horizontally and vertically around the saturated dayside of Venus and spreading with multiple reflections along the PtSi detector. Since this contamination is more reduced in the images taken with the 2.26-$\mathrm{\mu m}$ filter, we restricted our cloud tracking study to this data set composed of 466 images (although additional sets with 1.74-$\mathrm{\mu m}$ images were used in the case of automated cloud tracking). The calibration version of the IR2 images used in this work ("v20170601'') does not include any of the corrections for the light contamination proposed by \citet{Satoh2017} and an alternative image processing technique was applied. This consisted in an adjustment of the brightness/contrast, followed by sharpening the images with an unsharp-mask technique, and finishing with the application of adaptive histogram equalization (see examples in Fig.~\ref{figure:process-images}). Images acquired with ground-based instruments at TNG and IRTF also suffered from contamination from the illuminated side of the planet and were processed similarly. Some examples of these images are shown in section \ref{sec:TimeEvol}.\\

\subsection{Navigation of IR2 and ground-based images}\label{ssec:observ-navig}
\begin{figure*}
\centerline{\includegraphics[width=40pc]{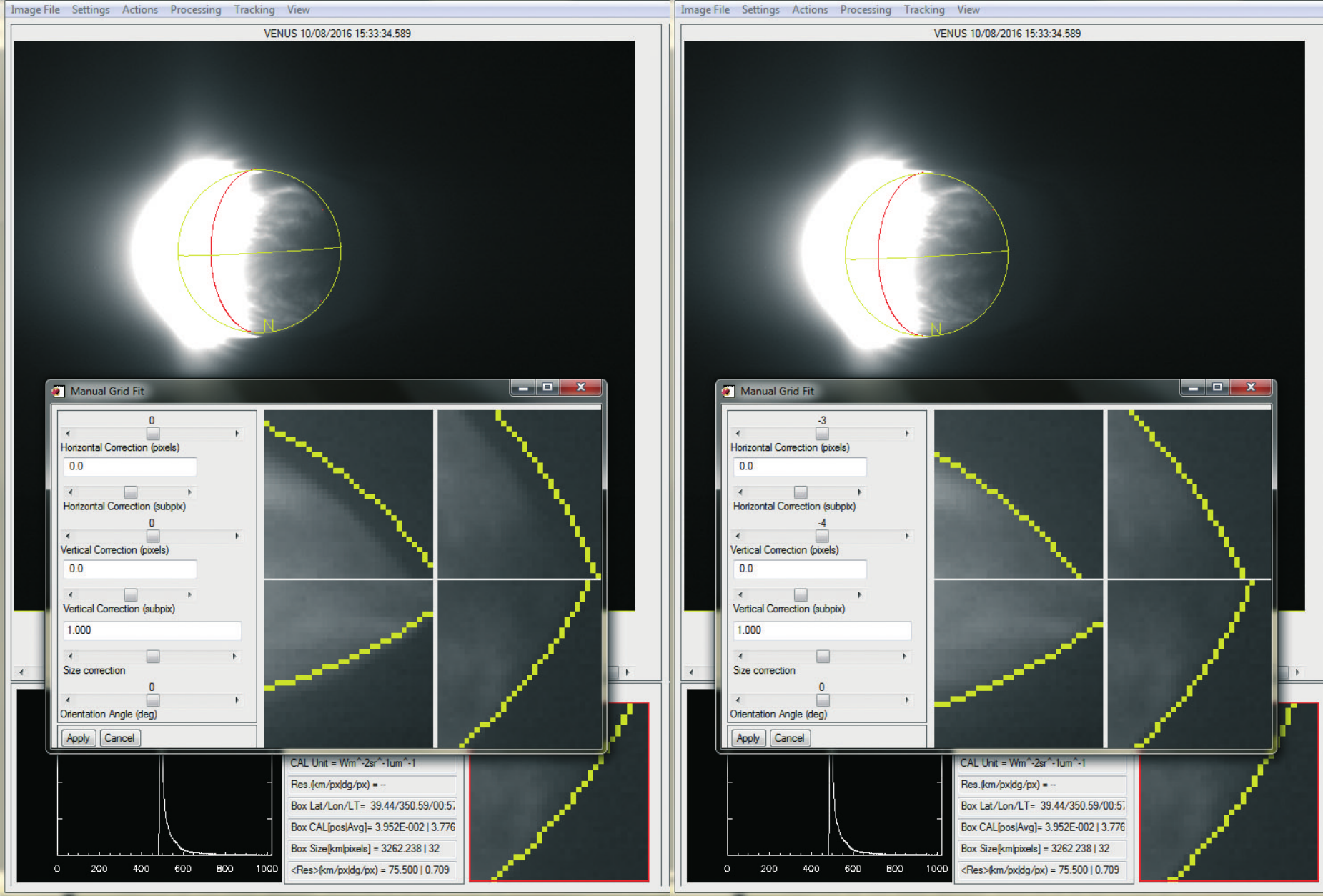}}
\caption{Example of the correction on the navigation of Akatsuki images using as reference a set of locations of the planetary limb selected by the user. The original navigation grid and the corrected one are displayed on the left and right sides, respectively.}
\label{figure:NavigCorrec}
\end{figure*}

Uncertainties such as the thermal distortion affecting the Akatsuki spacecraft and the ca\-me\-ras onboard prevent high accuracy in the navigation of the Venus images at present, so additional corrections in the navigation are yet required \citep{Ogohara2017,Satoh2017}. In recently published studies with Akatsuki \citep{Fukuhara2017NatGeo,Horinouchi2017NatGeo,Lee2017,Horinouchi2018} the navigation of the images was corrected with an algorithm able to perform an ellipse fitting from an automatic determination of the planetary limb pixels \citep{Ogohara2012,Ogohara2017}. This automatized identification of the limb becomes rather uncertain in many of the IR2 nightside images because of the light contamination previously described, and also due to the frequent darkening of the planetary limb because of the strong variability of the clouds' opacity at lower latitudes \citet{McGouldrick2008JGR,Satoh2017}. Instead, we coded an interactive tool inspired on the software WinJupos \citep{Hahn2012} which allows the interactive adjustment of the position, size and orientation of the planet's grid\footnote{The corrected geometry files are available from the corresponding author on reasonable request.}, using as reference four locations of the limb chosen by the user (see Fig.~\ref{figure:NavigCorrec}). The visualization of the limb was improved by interactively modifying the brightness and contrast in each image. In the case of the IR2 images, the position of the grid was adjusted with a precision of 1/10th of pixel, its size was increased in some cases (less than 1.3\% in all the cases), while the orientation of the grid required no corrections. Regarding the ground-based images from NICS and SpeX, these were navigated using NASA's SPICE kernels \citep{Acton1996,Folkner2009}, both the position and orientation of the grid were adjusted, and the predicted size was accurate enough to require no further corrections.\\

\subsection{Techniques for Cloud Tracking}\label{ssec:observ-track}
\begin{figure*}
%\centerline{\includegraphics[width=40pc]{f03.jpg}}
\centerline{\includegraphics[width=40pc]{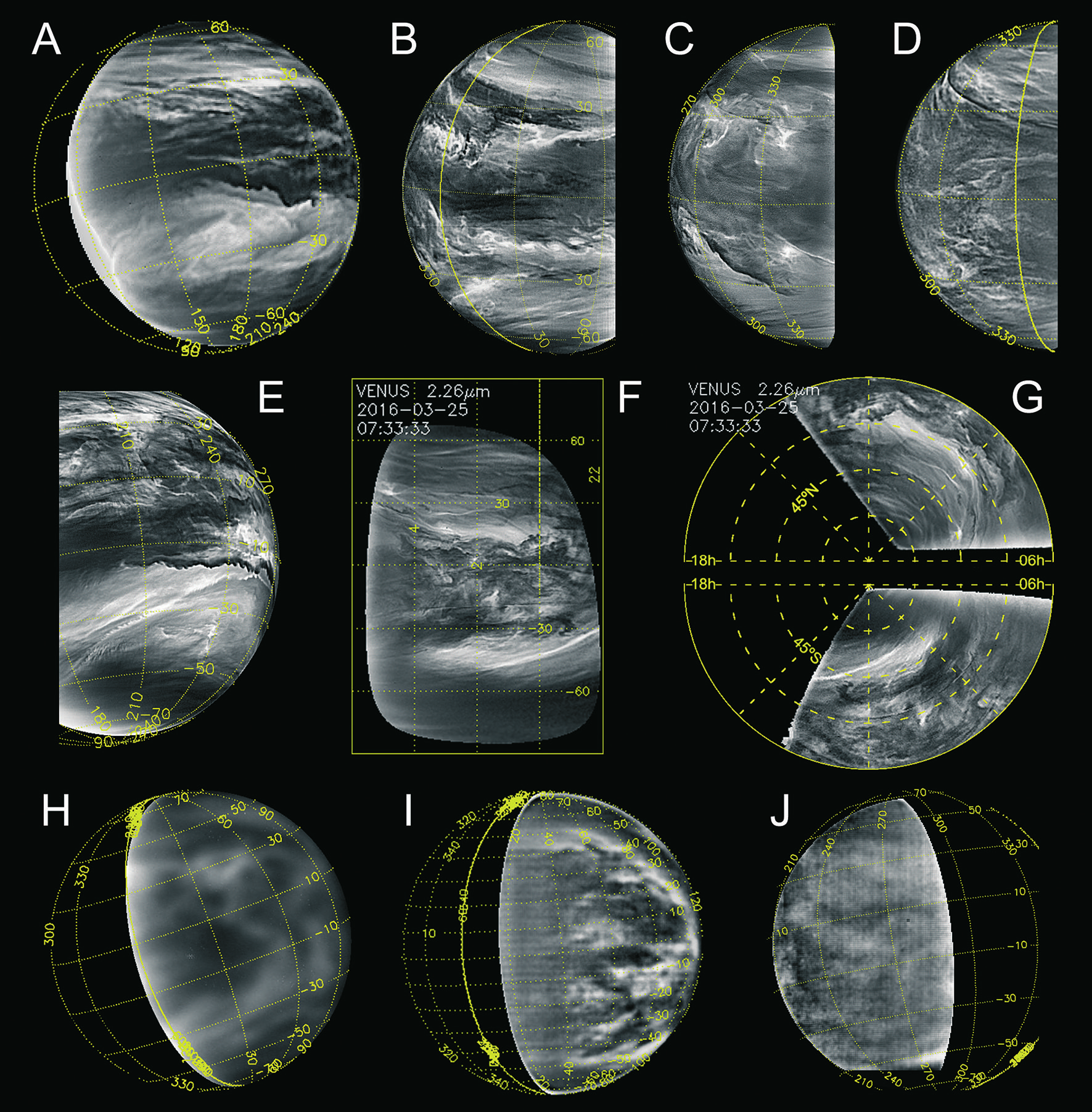}}
\caption{Examples of images acquired with the filter 2.26-$\mathrm{\mu m}$ by Akatsuki/IR2 during the year 2016 (\textbf{A--G}) and ground-based observations at 2.32-$\mathrm{\mu m}$ from TNG/NICS and IRTF/SpeX (\textbf{H--J}): (\textbf{A}) 2016 October 10, 09:23 UT, (\textbf{B}) 2016 August 13, 00:33 UT, (\textbf{C}) 2016 July 11, 22:03 UT, (\textbf{D}) 2016 July 22, 14:33 UT, (\textbf{E}) 2016 October 19, 14:33 UT, (\textbf{F}) cylindrical and (\textbf{G}) polar projections for 2016 March 25, 07:33 UT (original shown in Fig.~\ref{figure:distinctwinds}A), (\textbf{H}) 2012 July 11, 05:19 UT by TNG/NICS, (\textbf{I}) 2015 September 28, 16:03 UT by IRTF/SpeX, and (\textbf{J}) 2017 February 10, 02:31 UT by IRTF/SpeX. In the case of the polar projection (\textbf{G}), the northern and southern hemispheres are displayed above and below, respectively, with the geographical north/south pole located in the center.}
\label{figure:IR2images}
\end{figure*}

After correcting the navigation and processing the images, these were geometrically projected onto equirectangular (cylindrical) geometry with an angular resolution similar to the best resolution in the original images. In those cases of nightside images acquired when Akatsuki was closer to its pericenter, the better spatial resolution enabled measuring wind speeds at polar latitudes and azimuthal equidistant (polar) projections were performed too using an angular resolution similar to that found at latitudes of about 70$^{\circ}$ in the original image. Figure ~\ref{figure:IR2images} shows examples of the original observations, their navigation after corrections and cylindrical and polar maps. For the measurement of wind speeds, two different techniques of cloud tracking were employed: (a) a \textit{manual method} applied to the full data set of IR2 and ground-based images (see Table \ref{tab:IR2-observ}), consisting of a manual search of the cloud tracer followed by a fine adjustment using an automatic template matching which is visually accepted or rejected by the human operator (similarly as performed by \citealt{Hueso2015}); and (b) a \textit{fully-automatic method} used for images taken in 2016 July and August (see Table \ref{tab:IR2-observ}) and which applies the relaxation labeling technique \citep{Ikegawa2016,Horinouchi2017MST}.\\

\begin{figure}
%\centerline{\includegraphics[width=20pc]{f04.jpg}}
\centerline{\includegraphics[width=20pc]{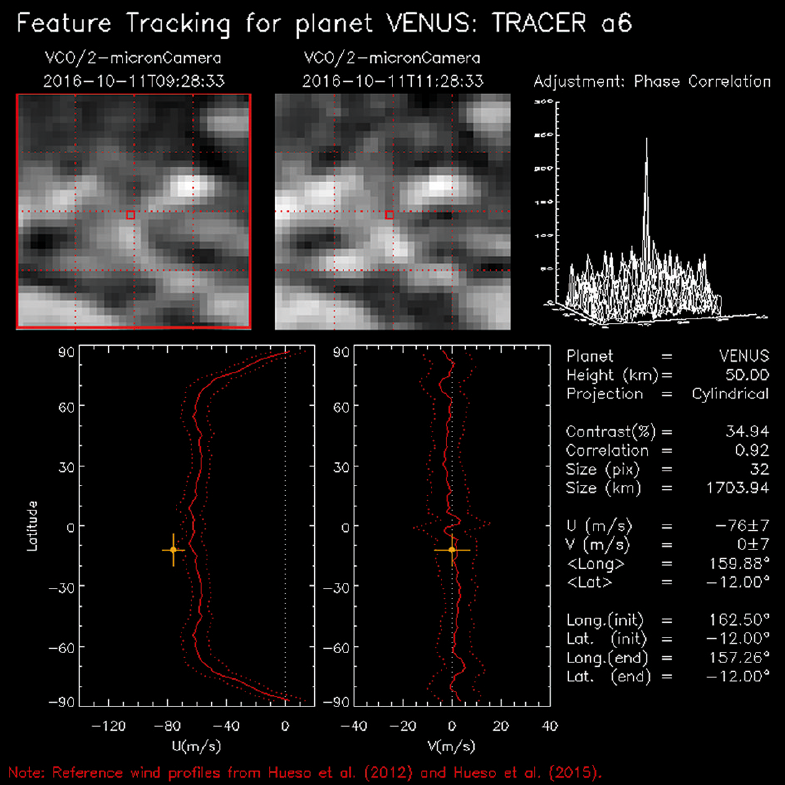}}
\caption{Example of positive identification of a cloud tracer in a pair of IR2 images using phase-correlation. Note that when a match is successful (see templates), a single and clear peak --ideally Kronecker delta-- is apparent in the cross-correlogram located in the corner right-up. The inferred wind speeds are compared with a reference wind profile from VEx/VIRTIS-M \citep{Hueso2012} in the two graphs below, while detailed information is provided on the right side below the correlogram. A complete report of clouds' matches with phase correlation can be found in the Supplemental Material (see Appendix \ref{sec:apdx}).}
\label{figure:PhaseCorrel-Example}
\end{figure}

Wind measurements are obtained by comparing the position on maps of cloud features that can be identified in two consecutive images with a given time difference. We obtained wind measurements with a \textit{manual method} in which we perform an automatic template matching with a \textit{phase-correlation} technique that has been also applied for cloud tracking on the Earth \citep{Leese1970,Jun1992,Humblot2005,Huang2012}. We call this technique \textit{manual} because a human user selects a region in the first image to obtain its best match in the second image and validates (or not) the final result. The phase-correlation permits to obtain the translation between two images shifted relative to each other \citep{Kuglin1975,Samritjiarapon2008}, relying on a frequency-domain representation of the images calculated with the Fast Fourier Transform that enables to infer this shift from the location of a peak in a \textit{cross-correlogram}. Figure~\ref{figure:PhaseCorrel-Example} shows an example of the use of this technique where the narrow peak in the phase cross-correlogram is shown for a well identified tracer. A more complete explanation about the phase-correlation is provided by \citet{Kuglin1976}. Compared to the standard correlation, its performance is faster, and it is less sensitive to random noise and illumination conditions \citep{Ahmed2008}. Even though the phase-correlation can potentially register sub-pixel displacements \citep{Reddy1996,Foroosh2002}, we did not test this capability in this work and only accounted for displacements expressed as integers of pixels. Besides, since the boundaries of any image imply discontinuities in the signal that can introduce a noisier result, a zero-padding is advisable before applying the FFT \citep{Ahmed2008}. For this reason, the initial and final templates (ranging sizes of 32$\times$32 to 48$\times$48 pixels) were convolved with a Hanning window with a width of 0.7 before the FFT was applied. When no clear peak could be found with the phase correlation, or when the tracer identification was judged as not satisfactory for the human operator running the analysis, the wind measurement was rejected and the template matching was performed manually. Each of these manual measurements was verified by a human operator looking at visual reports as the one shown in Figure~\ref{figure:PhaseCorrel-Example}.\\

A fully automatic technique based on classical image correlation for identifying the cloud tracers motions was also used by \citet{Horinouchi2017MST, Horinouchi2017NatGeo} and we will show here detail report of these measurements. This technique is similar to those used in many previous studies of Venus cloud dynamics  \citep{Rossow1990,Kouyama2012,Khatuntsev2017,Horinouchi2018}, but in this case the cross-correlation between images is computed along a sequence of images to estimate the horizontal velocity at a specific location \citep[see Methods therein]{Horinouchi2017NatGeo}. This method was applied to images of the lower clouds acquired in 2.26 and  1.74-$\mathrm{\mu m}$. All the IR2 images were projected onto equirectangular geometry with a fixed angular resolution of 0.125$^{\circ}$ per pixel regardless of the resolution of the original images \citep{Ogohara2017}. Image processing was applied after the projection and consisted on a two-dimensional bandpass filter with Gaussian functions with the sigma values 0.25$^{\circ}$ for low pass and 3$^{\circ}$ for high pass with latitude and longitude. Finally, the template size for cloud tracking was set to a fixed value of 60$\times$60 pixels for all the images.\\

\section{Meridional profiles of wind speeds}\label{sec:meridprofiles}
A total of 2,947 wind vectors were obtained with the manual method for the nightside images of the IR2 camera for the full period covering during 2016 from March 22 (when pairs of IR2 images were taken within the same day for the first time) until October 31. The spatial resolution of the images ranged 10--80 km per pixel and we selected pairs of images with time separations ranging from 1 hour up to 22 hours (in special cases of clouds displaying very small deformations over large time scales). The size of the cloud tracers comprise dimensions ranging 510--2,550 km, depending on the spatial resolution selected for the geometrical projections. Images acquired over 2016 April, July and August were also independently measured with the fully-automatic method described above resulting in 149,033 wind measurements from the tracking of cloud features with a fixed size of 790 km. The errors for the wind speeds obtained with manual tracking were calculated from the spatial resolution and the time interval between the images as explicated by \citet{Bevington1992}, while in the case of automatic cloud tracking the calculation of the errors are explained by \citet{Ikegawa2016}.\\
\\
\citet[fig.~11b therein]{Satoh2017} and \citet[fig.~15 therein]{Limaye2018} show Akatsuki IR2 images of strong cloud discontinuities in the lower clouds of Venus. These discontinuities propagate faster than the background zonal flow and were suspected to be the manifestation of waves rather than passive tracers. For this reason, we discarded the motions of these strong opacity discontinuities from our study of the global wind motions. In the case of the automatic cloud tracking, we ruled out all the measurements obtained from image pairs where this equatorial cloud discontinuity was apparent. As a result of this filtering, the number of wind vectors with the manual and automatic methods was reduced to 2,277 and 101,882 respectively. Since the automatic method does not provide a registry of the cloud morphologies tracked, this filtering also removed all automatic-generated wind vectors during April 2016. Table \ref{tab:IR2-observ} summarizes the results obtained with both techniques in different time periods and their coverage over the planet. The full report of wind measurements from the 2.26-$\mathrm{\mu m}$ IR2 images without filtering out specific features and obtained with the manual method can be found in the Supplemental Material (see Appendix \ref{sec:apdx}), including a table with the complete data set of wind measurements with manual tracking, geometrical projections, animations and detailed template matching results.\\

%\begin{deluxetable*}{ccCrlc}[b!]
\begin{deluxetable*}{ccccccc}
\tablecaption{Coverage of Akatsuki/IR2 Winds during year 2016}
\tablecolumns{6}
\tablenum{1}
\tablewidth{0pt}
\tablehead{
\colhead{Month} & \colhead{Dates\tablenotemark{*}} &
\colhead{Latitudes} & \colhead{Local Times} & \colhead{Longitudes} & \colhead{Wind Vectors\tablenotemark{**}}
}
\startdata
March & 22,23,25--30 & 61$^{\circ}$N--60$^{\circ}$S & 23h--05h & 225$^{\circ}$--308$^{\circ}$ & 249 \\
April & \textbf{15} & 43$^{\circ}$N--63$^{\circ}$S & 01h--06h & 281$^{\circ}$--348$^{\circ}$ & 35+\textbf{306} \\
June & 20 & 40$^{\circ}$N--21$^{\circ}$N & 18h--20h & 264$^{\circ}$--289$^{\circ}$ & 12 \\
July & 1,\textbf{11},12,22 & 72$^{\circ}$N--69$^{\circ}$S & 18h--01h & 250$^{\circ}$--23$^{\circ}$ & 308+\textbf{6920} \\
August & \textbf{2},9,\textbf{10},\textbf{13},\textbf{15--17},  & 58$^{\circ}$N--58$^{\circ}$S & 18h--03h & 322$^{\circ}$--129$^{\circ}$ & 1057+\textbf{94656} \\
       & 18,19,\textbf{20},21,22,\textbf{25}, &   &  &        &     \\
       & \textbf{26},27,28,\textbf{29},\textbf{30} &   &  &        &     \\
September & 4--6,15,26,27 & 59$^{\circ}$N--60$^{\circ}$S & 19h--05h & 28$^{\circ}$--188$^{\circ}$ & 347 \\
October & 2--7,10--17, & 56$^{\circ}$N--72$^{\circ}$S & 20h--05h & 114$^{\circ}$--270$^{\circ}$ & 939 \\
        & 19--27,31    &      &                        &          &     \\
\enddata
\tablenotetext{*}{Bold characters stand for dates with wind measurements from both manual and fully-automatic methods. Normal characters are used for dates with only manual measurements.}
\tablenotetext{**}{Normal and bold characters stand for number of wind measurements obtained with the manual and fully-automatic methods, respectively.}
%% \tablecomments{Description of wind speeds measured with the Akatsuki/IR2 images. In addition to the time coverage, spatial coverage is also specified in terms of Latitude, Local Time and Longitude.}
\label{tab:IR2-observ}
\end{deluxetable*}

\subsection{Morphologies of the nightside clouds}\label{ssec:observ-imgs}
Panels \textbf{A--G} in Figure \ref{figure:IR2images} exhibits a sample of the clouds' morphologies observed in the IR2/2.26-$\mathrm{\mu m}$ images. Pioneering ground-based observations at these wavelengths \citep{Allen1984,Crisp1991b} show similar global characteristics as Akatsuki IR2 images. The deeper clouds of Venus nightside are normally characterized by a dark band with high opacity clouds at low latitudes, while the mid-latitudes are dominated by brighter bands with lower opacity (see Figs.~\ref{figure:IR2images}\textbf{A--B}). Prior to Akatsuki, the mid-latitude bands appeared as almost featureless \citep{Crisp1991b,Limaye2006,Hueso2012}. However, the higher spatial resolution of the Akatsuki IR2 images reveals on them subtle though distinguishable wisps and patches (Figs.~\ref{figure:IR2images}\textbf{F--G}) sometimes invaded by broad bands of clouds with higher opacity (Fig.~\ref{figure:IR2images}\textbf{B}), or unusual sharp dark spirals tilted relative to the latitude parallels (Figs.~\ref{figure:IR2images}\textbf{C--D}) which spread thousands of kilometres from latitudes higher than 30$^{\circ}$ towards equatorial ones \citep{Horinouchi2017NatGeo,Satoh2017,Limaye2018}.\\
\\
In the IR2 images, the clouds' opacity at lower latitudes exhibits higher variability than at mid-latitudes \citep{Satoh2017,Limaye2018}, confirming previous findings with ground-based observations \citep{Crisp1991b,Limaye2006,Tavenner2008,Machado2016DPS} and during the VEx mission \citep{Hueso2012,McGouldrick2012,McGouldrick2017}. Dark areas of high opacity normally dominate lower latitudes, while their boundary with the mid-latitude bright bands exhibits complex cloud features \citep{Horinouchi2017NatGeo,Limaye2018}, like small vortices or abundant bright swirls at around 30$^{\circ}$ which sometimes adopt shapes suggestive of shear instabilities (Fig.~\ref{figure:IR2images}\textbf{B,C,E}), similar to those found at the nightside upper clouds at $\sim$65 km with 3.8-$\mathrm{\mu m}$ VEx images\citep{Peralta2017NatAstro}. The polar projection in Fig.~\ref{figure:IR2images}\textbf{G} clearly exhibits an example of the frequent episodes of hemispherical asymmetry for the Venus clouds' morphology. A more complete survey of the clouds' morphologies apparent in the images by the Akatsuki/IR2 camera will be presented elsewhere.\\

\subsection{Latitudinal profiles of the winds and relation with clouds}\label{ssec:relation-winds-clouds}
\begin{figure*}
\centerline{\includegraphics[width=45pc]{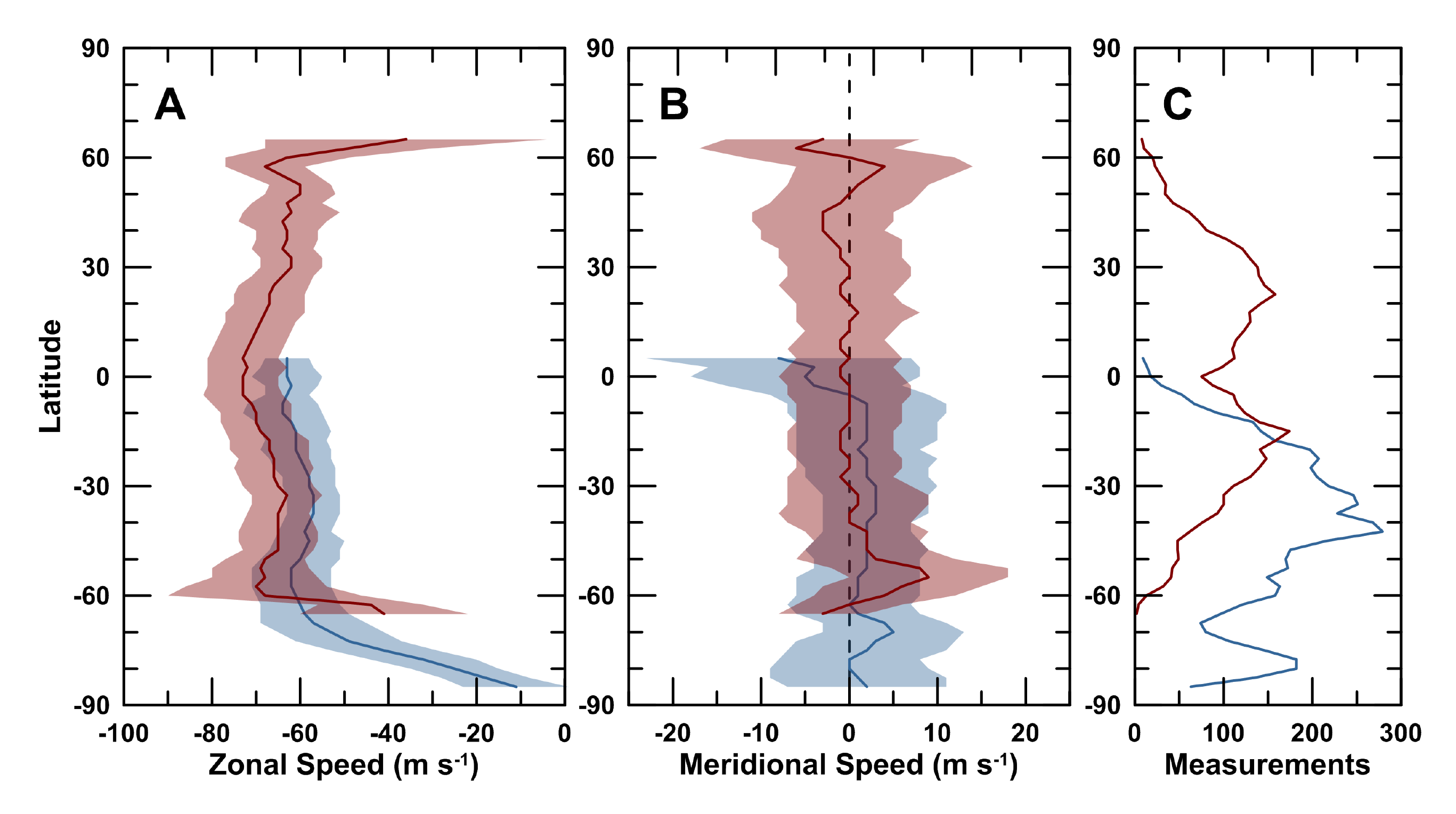}}
\caption{Mean profiles of the winds at the nocturnal middle-to-lower clouds of Venus as measured from March to October 2016 with the manual method using the Akatsuki/IR2 2.26-$\mathrm{\mu m}$ images (dark red line). These profiles are compared with the winds measured with 1.74-$\mathrm{\mu m}$ images acquired with VEx/VIRTIS-M (blue line) between 2006 April and 2008 August \citep{Hueso2012}. The profiles for the zonal and zonal and meridional components are shown in panels (\textbf{A}) and (\textbf{B}), respectively. Average values shown here were calculated for bins of latitude of 5$^{\circ}$, and the number of measurements used in each bin are displayed in panel \textbf{C}.The shadowed areas stand for the standard deviation at every latitude bin when this is larger than the measurement error computed from the image resolution and time differences in the image pair.}
\label{figure:windprofiles}
\end{figure*}

Figure~\ref{figure:windprofiles} shows zonally-averaged profiles of  the manually-tracked winds at the lower clouds during the first year of Akatsuki observations. Bins of 5$^{\circ}$ of latitude were considered to calculate average wind speeds from March to October 2016. The profiles of IR2 are compared with VEx/VIRTIS-M results from 2006 April to 2008 August \citep{Hueso2012}. With regards to the zonal component of the wind (Fig.~\ref{figure:windprofiles}\textbf{A}), the mean profile obtained with IR2 is faster than VIRTIS-M measurements by 10 ms$^{-1}$. The zonal winds from IR2 observations show symmetric profiles between both hemispheres and, despite the higher dispersion at higher latitudes due to the poorer spatial resolution of the IR2 images close to the poles, they show for the first time the decay of the winds towards the poles on both hemispheres simultaneously. This decay starts at about 60$^{\circ}$ and, compared to VEx results, it seems more abrupt, while it keeps similarities with wind profiles obtained in 2004 with ground-ground observations \citep[fig.~5 therein]{Limaye2006}, albeit, with higher errors. However, we cannot rule out that these differences at subpolar latitudes with respect to VEx/VIRTIS-M subpolar winds are not caused by the worse spatial resolution at polar latitudes and the low number of measurements in the Akatsuki IR2 images (Fig.~\ref{figure:windprofiles}\textbf{C}). Regarding the meridional component of the wind (Fig.~\ref{figure:windprofiles}\textbf{B}), the results of 2016 and 2006--2008 are in good agreement, confirming the absence of clear global motions in the meridional circulation of the nocturnal lower clouds.

\begin{figure}
%\centerline{\includegraphics[width=20pc]{f06.jpg}}
\centerline{\includegraphics[width=20pc]{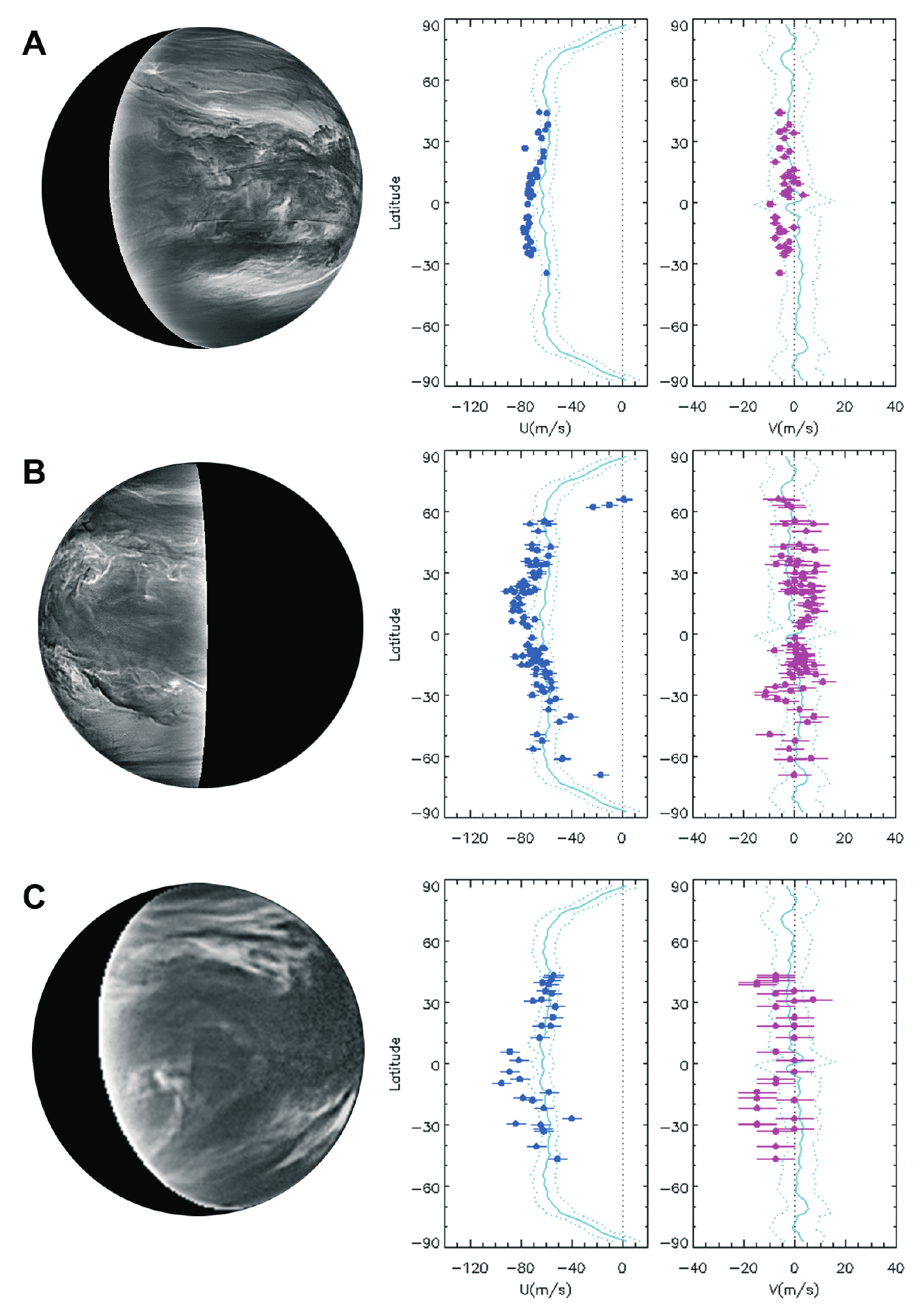}}
\caption{Variability in clouds' morphology (left column) and profiles of zonal and meridional winds (center and right column) during the Akatsuki mission in 2016: (\textbf{A}) winds during March 25, exhibiting the standard profile of constant zonal speeds; (\textbf{B}) winds during July 11, displaying a zonal jet at $\sim$20$^{\circ}$N; and (\textbf{C}) winds during October 13, exhibiting a strong jet at the equator and higher dispersion in the southern hemisphere. Reference profiles built from winds during the VEx mission \citep{Hueso2012} and plot in a symmetric way for the two hemispheres are displayed in cyan.}
\label{figure:distinctwinds}
\end{figure}

Wind variability is also apparent during 2016 (see Fig.~\ref{figure:distinctwinds}). The profile obtained in 2016 March 25 (Fig.~\ref{figure:distinctwinds}\textbf{A}) corresponds to winds that are approximately constant between the equator and mid-latitudes with small meridional shear. This seems to be most frequent case found at the lower clouds during the Akatsuki observations and also during the VEx mission \citep[fig.~7 therein]{Hueso2012}. Zonal wind profiles on other dates like the July 11 (Figs.~\ref{figure:distinctwinds}\textbf{B}) and October 13 (\ref{figure:distinctwinds}\textbf{C}) exhibit more intense equatorial zonal speeds, first-time reported during the Galileo flyby \citep[fig.~4 therein]{Crisp1991b} and later identified as recurrent episodes of jets at the equator \citep{Horinouchi2017NatGeo}. There is a reasonable correspondence between this local intensification of the zonal speeds and the presence of features resembling shear instabilities (Fig.~\ref{figure:distinctwinds}\textbf{B}) or sharp opacity discontinuities at the equator (Fig.~\ref{figure:distinctwinds}\textbf{C}). The high dispersion in the horizontal speeds also suggests that these jets may be apparent only in a rather longitudinally narrow area. The profile during July 11 (Fig.~\ref{figure:distinctwinds}\textbf{B}) shows that these jets can sometimes show up at northern latitudes, as also reported in 1990 during the Galileo flyby \citep[fig.~6 therein]{Carlson1991} and in 1996 from observations at the Apache Point Observatory \citep[fig.~7 therein]{Chanover1998}.\\

\section{Sources of variability for the wind speeds at the lower clouds.}\label{sec:WindDependences}
In this section, we explore whether the size of the features tracked,  the opacity of the clouds, the uncertainties in their vertical sensing, or a dependence on the local time and surface elevations, can help to explain the variability observed in the winds of the nightside lower clouds. 

\subsection{Effect of opacity and size of the clouds}\label{ssec:Rad-Size}
\begin{figure}
\centerline{\includegraphics[width=20pc]{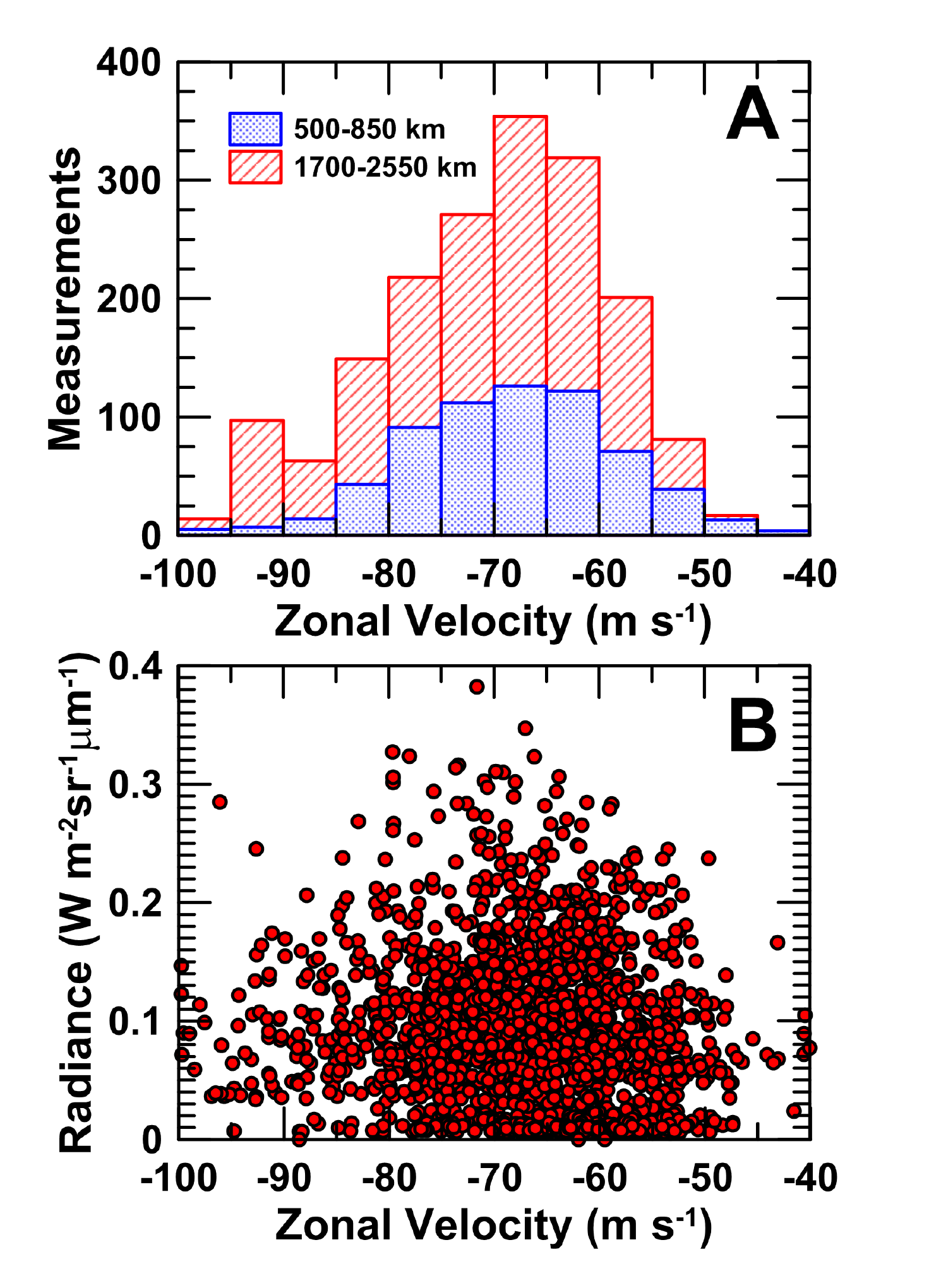}}
\caption{Histograms and scatter plot of the wind measurements acquired with the manual method and Akatsuki/IR2 images. The panel \textbf{(A)} shows histograms of the zonal velocities for two ranges of spatial scales of cloud tracers. The panel \textbf{(B)} exhibits the values of zonal wind speeds in terms of the mean radiance (red dots). The radiance is given for the calibration version "v20170601'' of the IR2 2.26-$\mathrm{\mu m}$ images, and corresponds to the radiance averaged inside the template.}
\label{figure:Rad-Size-Dep}
\end{figure}

It is yet unclear whether the episodes of faster zonal speeds shown in Fig.~\ref{figure:distinctwinds} may be related to horizontal and/or vertical gradients of the zonal wind, or if the bright and dark clouds observed on the nightside at 1.74, 2.26 and 2.32 $\mathrm{\mu m}$ are moving at the same or at slightly different vertical levels. Early interpretations of Venus IR features considered that these clouds' contrasts might be caused either by scattered sunlight leaking in from the dayside along altitudes with low absorption in the CO$_{2}$ windows, or that dark areas corresponded to opaque clouds lying in a broken layer at certain altitude above the brighter regions and radiate as blackbodies at the lower temperature of their tops \citep{Allen1984}. \citet{Allen1987} discarded both interpretations, confirming that the existence of important differences in the vertical elevation of bright and dark clouds was inconsistent with the small variations that dark/bright areas exhibit in CO absorptions affecting the wide 2.2-2.3 $\mu$m band. Therefore, the dark and bright regions should correspond to horizontal inhomogeneities in the clouds' opacity to the deeper background thermal emission \citep{Allen1987,Crisp1989}, although certain discrepancies of altitude between dark and bright clouds cannot be discarded since these inhomogeneities can be caused by variations in the size and distribution of the particles within the lower and middle clouds \citep{McGouldrick2008Icarus}.\\
\\
Assuming a range of 50--60 km for the altitudes sensed at the relevant wavelengths \citep{McGouldrick2008JGR}, a comparison with \textit{in situ} wind profiles from descending probes \citep{Gierasch1997,Counselman1980,Moroz1997} indicates that variations of up to 30 m s$^{-1}$ could be explained in terms of the vertical shear \citep[fig.~3 therein]{Peralta2017NatAstro}, what is also consistent with the magnitude of the reported jets \citep[fig.~2b therein]{Horinouchi2017NatGeo}. \citet{Crisp1991b} obtained a discrepancy between velocities of large dark clouds and the smaller markings (sizes ranging 400--1000 km), and also suggested that these might be produced at different altitudes. Figure \ref{figure:Rad-Size-Dep} displays the distribution of zonal speeds obtained with the manual method compared with the averaged radiance (Fig.~\ref{figure:Rad-Size-Dep}\textbf{B}) and the size of the cloud tracers between 50$^{\circ}$N--50$^{\circ}$S (Fig.~\ref{figure:Rad-Size-Dep}\textbf{A}) in the IR2 2.26-$\mathrm{\mu m}$ images. No dependence is apparent between the zonal velocities and these parameters. Since the low latitudes frequently exhibit cloud patterns that resemble shear instabilities for a wide range of scales (see Figs.~\ref{figure:IR2images}\textbf{C--E}; \citealt[fig.~10 therein]{Limaye2018}), the generation of jets due to meridional gradients in the horizontal winds seems probable. \citet{McGouldrick2007,McGouldrick2008Icarus} showed that large-scale dynamics can also explain the strong variations in the cloud opacity, and that even weak downwelling is able to produce optical-depth holes in the clouds.\\

\subsection{Local time dependence.}\label{ssec:LTimeDep}
\begin{figure*}
%\centerline{\includegraphics[width=40pc]{f08.jpg}}
\centerline{\includegraphics[width=45pc]{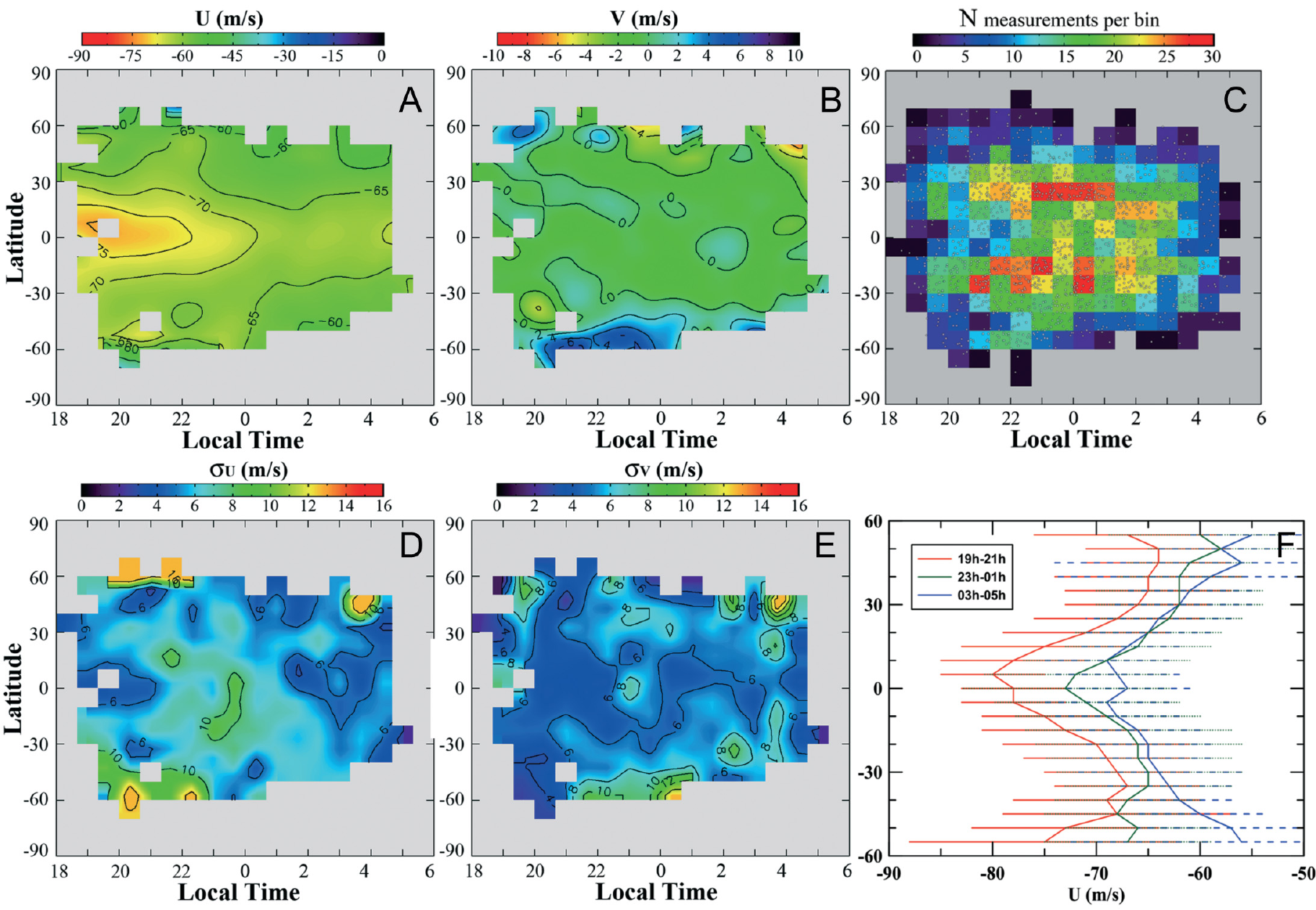}}
\caption{Local time dependence for the zonal winds obtained with the manual method on the Akatsuki/IR2-2.26-$\mathrm{\mu m}$ images. Bins of 10$^{\circ}$ in latitude and 40 minutes in local time were used. Panels (\textbf{A}) and (\textbf{B}) display the values of zonal and meridional components of the wind averaged in each bin; panel (\textbf{C}) shows the distribution of wind measurements with dots representing individual measurements; (\textbf{D}) and (\textbf{E}) display the standard deviation of the measurements in each bin; and (\textbf{F}) shows the meridional profiles of zonal winds at different intervals of local time.}
\label{figure:DependLTime}
\end{figure*}

Since most of the sunlight absorption occurs on Venus within the clouds' layer, solar tides are expected to be excited and carry momentum away upwards and downwards from the region of excitation, thus accelerating the atmosphere westwards and contributing to the RZS \citep{Gierasch1997,Sanchez-Lavega2017}. The effect of the solar tides have been unambiguously detected on both zonal and meridional components of the winds at the upper clouds at 65--70 km \citep{Rossow1990,Limaye2007,Sanchez-Lavega2008,Kouyama2012,Peralta2012,Khatuntsev2013,Hueso2015}, but no evident influence has been found for the middle-to-lower clouds (50--60 km) for either day \citep{Hueso2015,Khatuntsev2017} or night \citep[fig.~6 therein]{Hueso2012}. \citet{Khatuntsev2017} argued that this negligible effect of the solar tides could be related to the absence of the unknown absorber downward of the middle clouds (which is responsible for most of the solar heat deposition on Venus).\\
\\
Figure \ref{figure:DependLTime} allows to study the local time dependence for the winds at the nightside lower clouds of Venus during 2016. In agreement with VEx results during 2006--2008, no local time dependence is apparent in the meridional winds (Fig.~\ref{figure:DependLTime}\textbf{B}). However, the zonal component of the wind speed displays a local increase of the zonal speeds between 19--22 LT, followed by a gradual decrease towards late nightside local times (decrease of $\sim$10 m s$^{-1}$ between 19 LT and midnight). \citet{Hueso2012} also reported hints of faster retrograde winds close to dawn during 2006--2008, although this local time effect was weaker and disregarded due to the small number of measurements in this area. Recent results from Venus General Circulation Models (h.a. GCMs) predict that the diurnal tide should be also apparent on the zonal wind down to the middle clouds at $\sim$60 km with the slowest speeds centered at the evening terminator \citep[fig.~3a therein]{Takagi2018}. Better agreement is found for the GCM results at the lower clouds ($\sim$50 km), where the influence of the diurnal tide is predicted to weaken but a local maximum of about 10 m s$^{-1}$ is also found in the early night and before the midnight \citep[fig.~4a therein]{Takagi2018}.\\

\subsection{Longitudinal dependence.}\label{ssec:LongDep}
\begin{figure*}
%\centerline{\includegraphics[width=40pc]{f09.jpg}}
\centerline{\includegraphics[width=45pc]{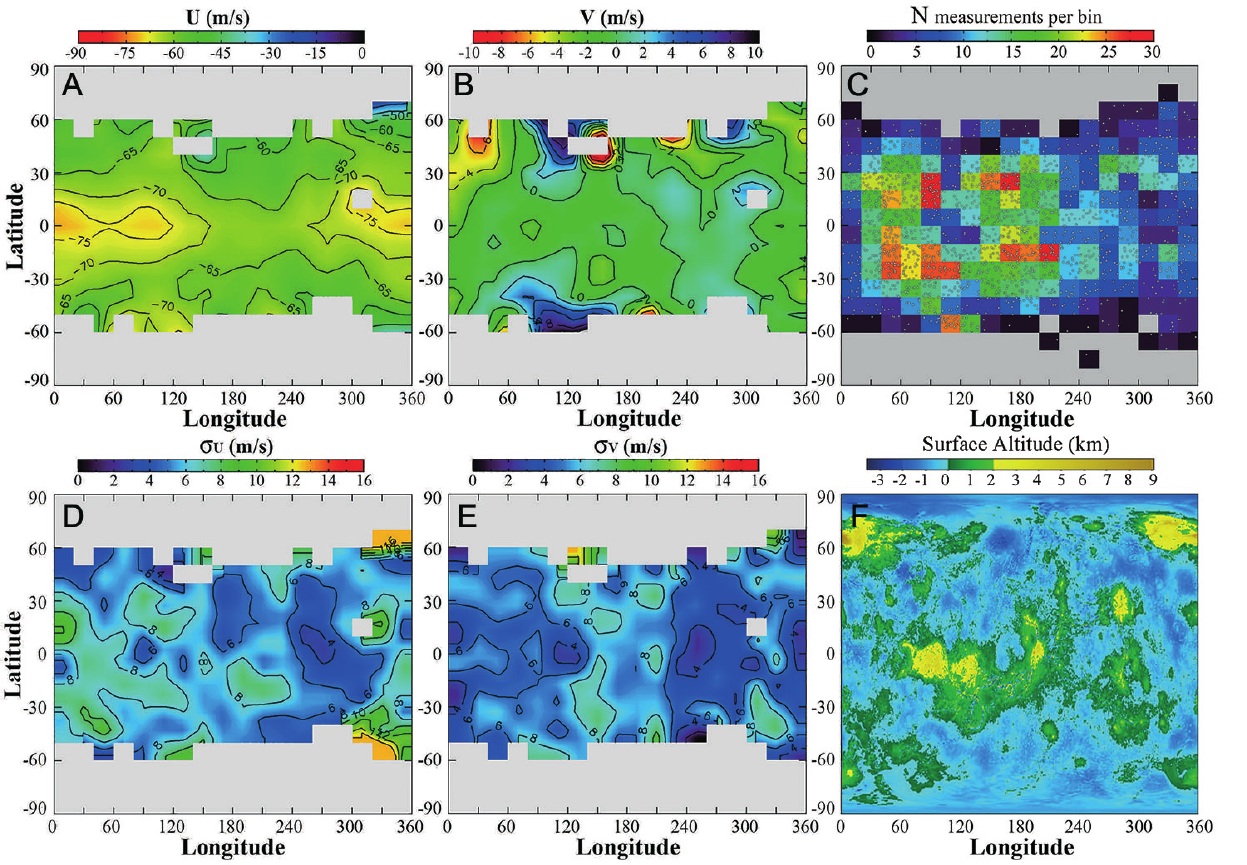}}
\caption{Local time dependence for the zonal winds obtained with the manual method on the Akatsuki/IR2-2.26-$\mathrm{\mu m}$ images. Bins of 10$^{\circ}$ and 40' were considered for the latitude and local time, respectively. Panels (\textbf{A}) and (\textbf{B}) display the values of zonal and meridional components of the wind averaged in each bin; panel (\textbf{C}) shows the distribution of wind measurements; (\textbf{D}) and (\textbf{E}) display the error (standard deviation) in each bin; and (\textbf{F}) exhibits the surface elevation of Venus as a comparison.}
\label{figure:DependLong}
\end{figure*}

Evidences on how the surface topography may be influencing the atmospheric circulation through the excitation of atmospheric stationary waves (lee waves) have been accumulating. Vertical disturbances experienced by the VEGA-2 balloon over Aphrodite terra \citep{Blamont1986}, strong asymmetries of the water vapor over certain geographical locations \citep{Fedorova2016}, or the finding of multiple stationary waves at the upper clouds of Venus which are strongly correlated with the surface elevations \citep{Peralta2017NatAstro,Fukuhara2017NatGeo,Kouyama2017} strongly suggest surface effects on the upper atmosphere. Results of cloud tracking in VEx/VMC dayside images seem to support that the wind speeds at the cloud tops are decelerated as they pass over Aphrodite terra and Atla regio \citep[figs.~4 and 6 therein]{Bertaux2016}. Other works have reported indications of an effect of surface elevations for the winds at the middle cloud \citep[fig.~14 therein]{Khatuntsev2017} and over the oxygen airglow patterns \citep{Gorinov2018}.\\
\\
\citet{Peralta2017NatAstro} discovered that during 2006--2008 the RZS at the nocturnal upper clouds exhibited a higher variability compared to the dayside, but the geographical coverage of the wind measurements was insufficient to confirm a correlated effect with the surface elevations. In the case of the lower clouds on the nightside, stationary waves are paradoxically missing in the clouds' opacity and no longitudinal dependence had been reported for the wind speeds (a reanalysis of VIRTIS-M winds for the lower clouds resulted inconclusive due to the lack of enough data to separate local time effects from time variability and the elusive surface dependence). Figure \ref{figure:DependLong} displays the dependence on the longitude and latitude for the manual wind measurements with the Akatsuki IR2 images. While no clear effect is found for the meridional component of the wind (Fig.~\ref{figure:DependLong}\textbf{B}), the westward windspeeds exhibit a local maximum at low latitudes over longitudes ranging 300$^{\circ}$--120$^{\circ}$ (fig.~\ref{figure:DependLong}\textbf{A}).\\

\begin{figure}
%\centerline{\includegraphics[width=20pc]{f10.jpg}}
\centerline{\includegraphics[width=20pc]{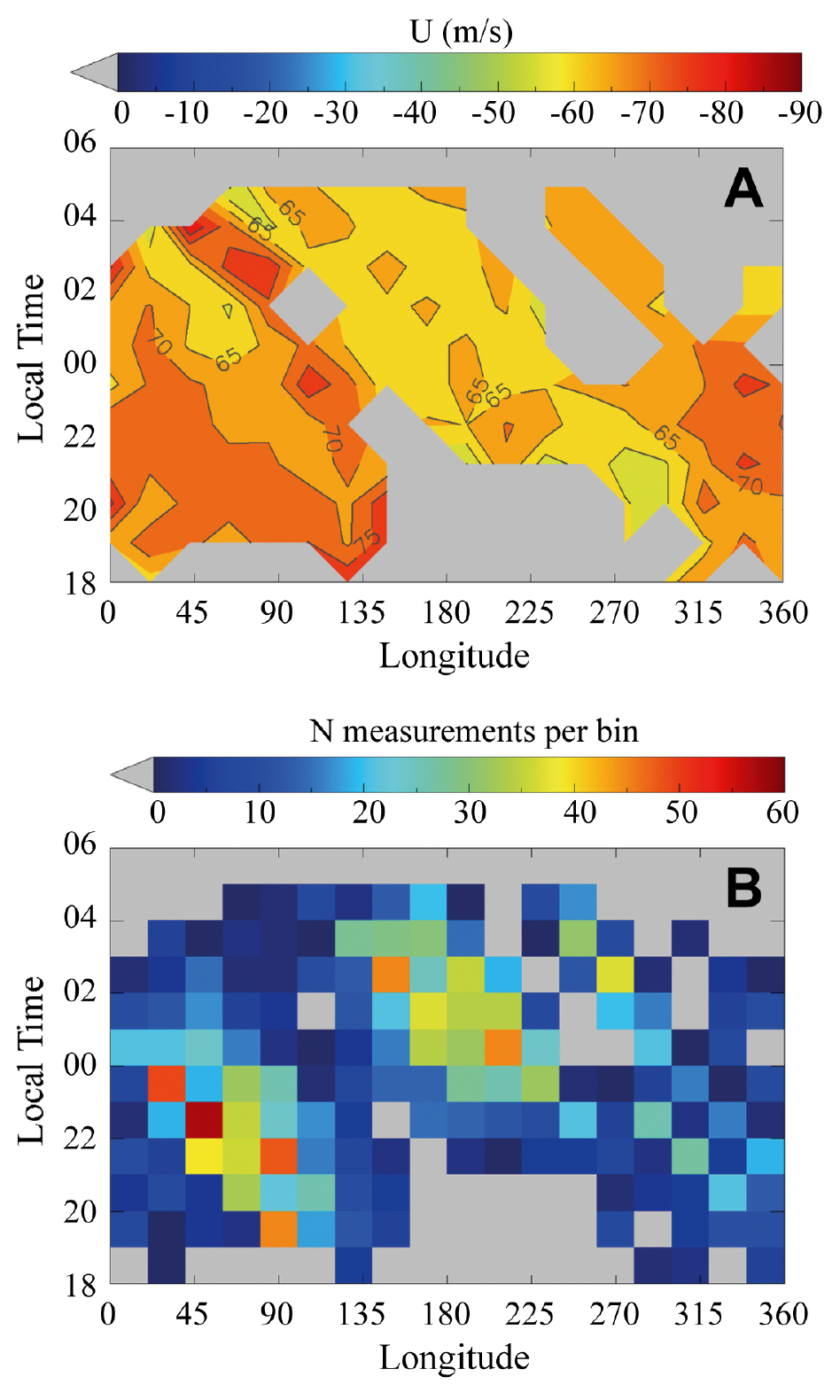}}
\caption{Distribution as a function of Longitude and Local Time of the wind measurements acquired with the manual method and Akatsuki/IR2 images between 50$^{\circ}$N--50$^{\circ}$S. The panel \textbf{(A)} displays the distribution of the zonal wind speeds. Panel \textbf{(B)} shows the distribution of the number of measurements.}
\label{figure:Dist-Long-Ltime}
\end{figure}

Unfortunately, separating the effects of the longitudinal and local time dependences is not feasible with our results, since the observed velocity disturbances are just marginally larger than the corresponding errors and the incomplete coverage of our wind measurements. Figure \ref{figure:Dist-Long-Ltime} shows the distribution of the mean zonal speeds (panel \ref{figure:Dist-Long-Ltime}\textbf{A}) and the number of measurements (\ref{figure:Dist-Long-Ltime}\textbf{B}) in terms of local time and longitude. To avoid the expected decrease of the zonal winds towards the poles, only the Akatsuki/IR2 wind measurements between 50$^{\circ}$N--50$^{\circ}$S were considered. Panel \ref{figure:Dist-Long-Ltime}\textbf{B} shows that the number of wind measurements during the year 2016 is irregularly distributed for both longitude and local time parameters, preventing a confirmation of either a local time dependence and/or influence of surface elevations.\\

However, the influence of the surface elevations over the zonal winds at the nightside lower clouds might be yet regarded as controversial. Published results for the zonal winds on the dayside were originated from a huge set of wind measurements with noticeable dispersion \citep[fig.~2 therein]{Bertaux2016}, and authors do not report filtering out other important sources of variability such as transient waves or the local time dependence. In the case of zonal wind speeds obtained with Akatsuki/UVI images, \citet{Horinouchi2018} characterized and removed the local time dependence on the zonal winds at the dayside upper clouds. As a result, no effect of surface elevations was found for the dayside winds \citep[fig.~14b therein]{Horinouchi2018}. In the case of our results for the nightside lower clouds, the local maximum extends mostly over rather plain areas and do not seem well correlated with the surface elevations at lower latitudes. Moreover, our results seem inconsistent with those on the dayside reported by \citet{Bertaux2016} and \citet{Khatuntsev2017} since, conversely to the zonal wind speeds on the dayside from VEx/VMC which exhibit a local minimum with slower speeds extending along 0$^{\circ}$--200$^{\circ}$ \citep[fig.~3 therein]{Bertaux2016}, our westward winds on the nightside display a local maximum between longitudes 300$^{\circ}$--120$^{\circ}$.\\

\subsection{Results from automatically retrieved winds during 2016 July and August}\label{ssec:ResultsAuto}
\begin{figure*}
%\centerline{\includegraphics[width=40pc]{f11.jpg}}
\centerline{\includegraphics[width=45pc]{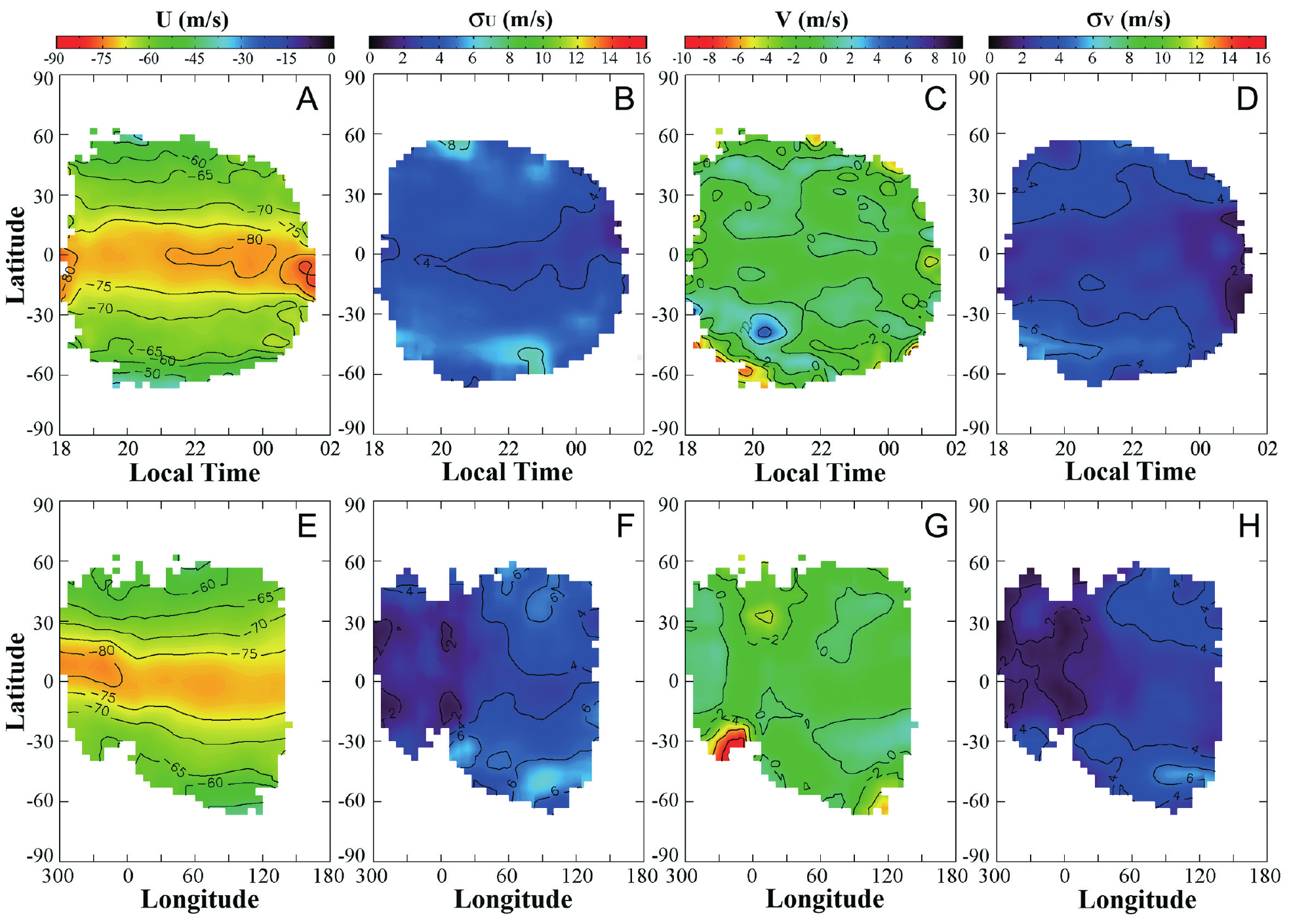}}
\caption{Contour maps for the dependence of zonal and meridional winds obtained with the full-automatic method in the Akatsuki/IR2 images. Averages (panels \textbf{A},\textbf{C},\textbf{E},\textbf{G}) and standard deviation (\textbf{B},\textbf{D},\textbf{F},\textbf{H}).The larger number of wind measurements allows to calculate these maps with bins of latitude, longitude and local time of 3.3$^{\circ}$, 3.3$^{\circ}$ and 13 minutes respectively, or three times smaller than in the manual tracking results. However, the spatial coverage over local times and longitudes is more limited in this case due to the data coming from a smaller number of dates.}
\label{figure:DependAutoWinds}
\end{figure*}

We now explore possible winds dependencies with the factors shown above for our wind results obtained in the automatic cloud correlation technique and firstly reported by \citet{Horinouchi2017NatGeo}. The dependence with both local time and longitude for the winds obtained with the fully-automatic method are displayed in Figure \ref{figure:DependAutoWinds}. Since for these wind speeds only IR2 images obtained during 2016 July and August were used (see Table \ref{tab:IR2-observ}), the coverage is more limited than in the case of the manual method, and the dispersion is smaller than for the manual results (see figs.~\ref{figure:DependLTime}\textbf{D},\ref{figure:DependLTime}\textbf{E},\ref{figure:DependLong}\textbf{D} and \ref{figure:DependLong}\textbf{E}).\\

Except for the case of the meridional winds where no clear dependence with the local time and longitude is found, the zonal winds with automatic correlation are not fully consistent with that from manual wind measurements, and no influence over the zonal speeds is observed at any range of local time or longitude. Even though some discrepancies were expected between both cloud tracking methods and the different techniques to correct the navigation of the IR2 images, the homogeneity found for the zonal winds at low latitudes might be explained due to the recurrence of the equatorial jet during August 2016 \citep{Horinouchi2017NatGeo}, which would mask a weaker dependence on local time and/or longitude.\\

\subsection{Combined results of Akatsuki and Venus Express}\label{ssec:CombVCOVEx}
\begin{figure*}
%\centerline{\includegraphics[width=40pc]{f12.jpg}}
\centerline{\includegraphics[width=45pc]{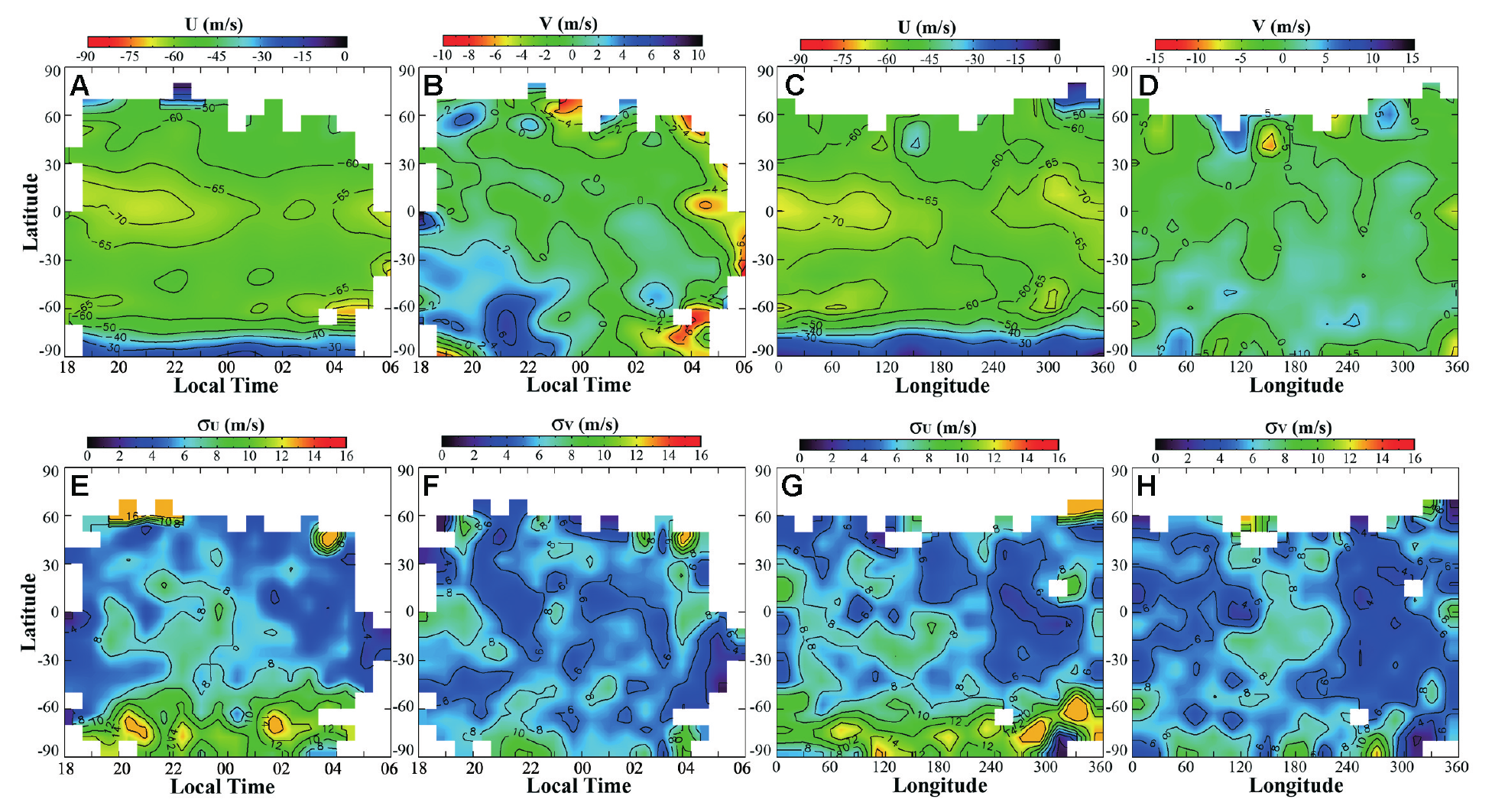}}
\caption{Winds' dependence with local time and longitude when the measurements from VEx/VIRTIS-M (2006--2008) and Akatsuki/IR2 (2016) are combined. Averages and standard deviation are calculated for bins of latitude, longitude and local time of 10$^{\circ}$, 10$^{\circ}$ and 40' respectively. Panels (\textbf{A}--\textbf{D}) display the zonal and meridional winds' dependence with the local time and the longitude, while panels (\textbf{E}--\textbf{H}) display the corresponding errors. Since the mean zonal winds during 2016 were faster than during the VEx mission (see Fig.~\ref{figure:windprofiles}\textbf{A}), the zonal speeds used in panel (\textbf{A}) and (\textbf{C}) where multiplied by a correction factor to have comparable zonally averaged profiles.}
\label{figure:DependLTimeLongVIRTIS}
\end{figure*}

We also combined the manual wind speeds at the nocturnal lower clouds with Akatsuki/IR2 in the year 2016 with those obtained by \citet{Hueso2012} with VEx/VIRTIS-M from 2006 to 2008. Since the mean zonal winds during 2016 were faster than during the VEx mission (see Fig.~\ref{figure:windprofiles}\textbf{A}), the difference between the zonal averages of both datasets equatorward of midlatitudes was suppressed by multiplying by a correcting factor of 1.06 the zonal speeds obtained with the VEx/VIRTIS-M images, while those from Akatsuki/IR2 images were divided by 1.06. As a result, a correction of $\sim$4 m s$^{-1}$ was introduced in both sets of winds. Also, these two datasets have a comparable number of points and both are based on cloud tracking inspected/validated by human operators. The result of their combination is presented in the figure \ref{figure:DependLTimeLongVIRTIS}. The dependence with the local time (panels \ref{figure:DependLTimeLongVIRTIS}\textbf{A--B, E--F}) confirms the equatorial maximum of the zonal speeds between local hours 18h--22h, while this dependence clearly decays towards higher latitudes. On the other hand, the meridional component of the wind displays no apparent dependence at low latitudes, while at higher latitudes of the southern hemisphere (50$^{\circ}$S--90$^{\circ}$S) and between 21h--22h it exhibits an equatorward acceleration (panel \ref{figure:DependLTimeLongVIRTIS}\textbf{B}). These results, again, suggest that the solar tide detected on the winds of the upper clouds \citep{Limaye2007,Peralta2012} might be able to propagate downwards to the middle clouds as predicted by some Venus GCMs \citep{Takagi2018}. Regarding the dependence with longitude and surface elevations, the local maximum between longitudes 300$^{\circ}$--120$^{\circ}$ reported in subsection \ref{ssec:LongDep} for the westward windspeeds is observed (panel \ref{figure:DependLTimeLongVIRTIS}\textbf{C}), while the meridional component exhibits no clear influence from the surface (panel \ref{figure:DependLTimeLongVIRTIS}\textbf{D}).\\

\section{Time Evolution of the Winds}\label{sec:TimeEvol}
\begin{figure*}
%\centerline{\includegraphics[width=40pc]{f13.jpg}}
\centerline{\includegraphics[width=45pc]{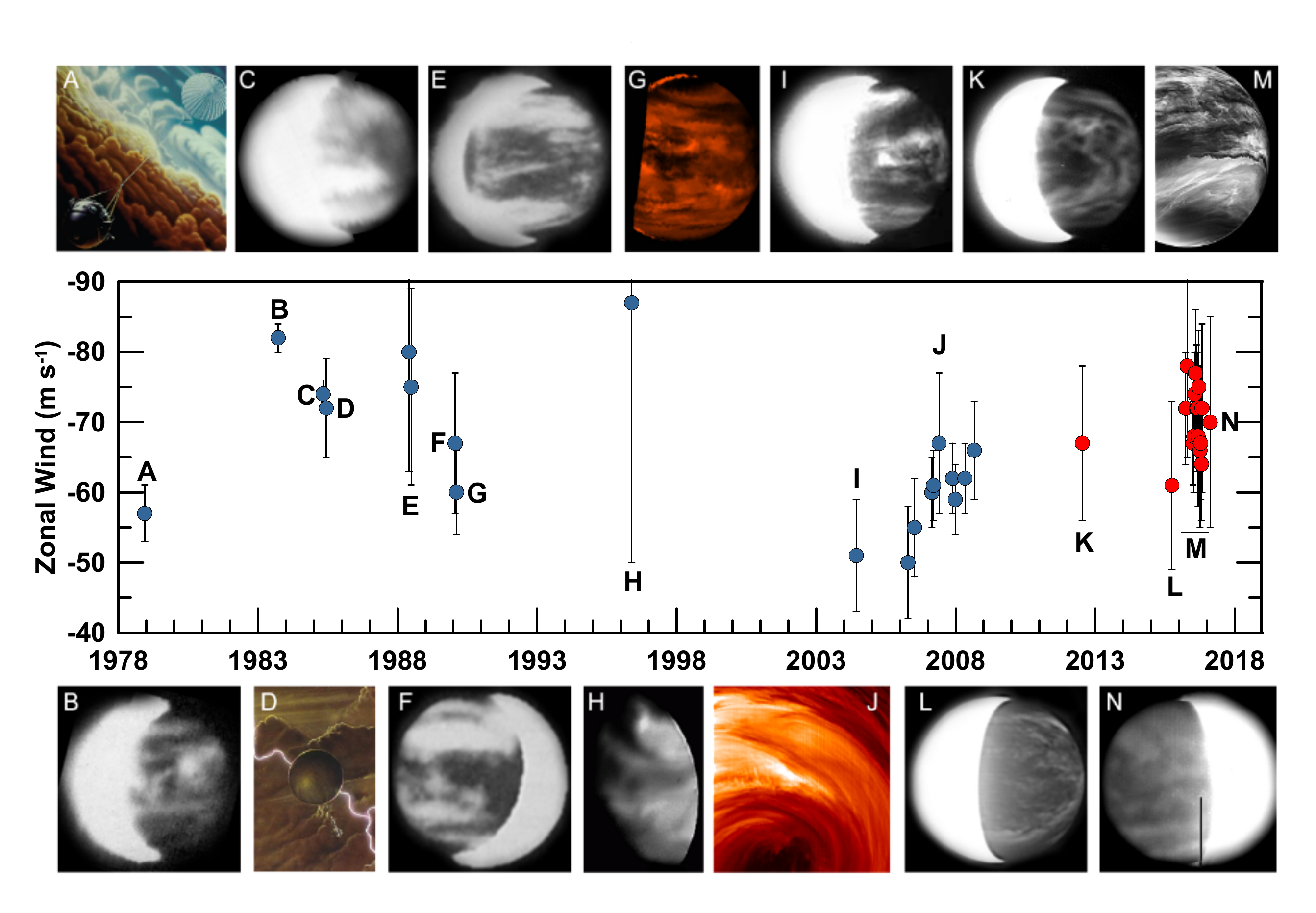}}
\caption{Decadal variation of the zonal winds at the nightside lower clouds of Venus. Data correspond to time-averages of the zonal winds obtained between 30$^{\circ}$S--30$^{\circ}$N with cloud tracking, except for \textbf{(A)} and \textbf{(D)} that represent instantaneous \textit{in situ} wind measurements from the Pioneer Venus Night probe \citep{Counselman1980} and VEGA landers \citep{Moroz1997} averaged within 50--60 km of altitude. Blue dots represent time-averages using wind speeds provided in past publications: \textbf{(A)} \citet{Counselman1980}, \textbf{(B)} \citet{Allen1984}, \textbf{(C)} \citet{Allen1987}, \textbf{(D)} \citet{Moroz1997}, \textbf{(E)} \citet{Crisp1989},  \textbf{(F)} \citet{Crisp1991b}, \textbf{(G)} \citet{Carlson1991}, \textbf{(H)} \citet{Chanover1998}, \textbf{(I)} \citet{Limaye2006}, and \textbf{(J)} \citet{Hueso2012}. New data presented in this work and based in cloud tracking measurements are displayed with red dots and were obtained from TNG/NICS \textbf{(K)}, IRTF/SpeX \textbf{(L and N)}, and Akatsuki/IR2 images \textbf{(M)}. The error bars stand for the standard deviation of the time averages.}
\label{figure:DecadalWinds}
\end{figure*}

Results from past space missions to Venus and comparison between them have provided evidence of long-term variability affecting the atmospheric dynamics at the upper clouds of Venus \citep[fig.~2 therein]{Sanchez-Lavega2017}. \citet{Rossow1990} reported that during the Pioneer Venus zonal winds at equator and mid-latitudes exhibited variations ranging 5--8 m s$^{-1}$ over time spanning 1--6 years, while the intensity of the poleward circulation seemed also subject to variability. From the analysis of the ultraviolet albedo of the cloud tops in different missions, \citet{delGenio1990} suggested that between 1979 and 1986 the clouds' dynamics might experience cyclic changes with a time scale of 5--10 years. This conclusion was based on the periods of presence/absence of the 4 and 5-day wave modes, while during the VEx mission \citet[fig.~18 therein]{Lee2015Icar} reported unexplained periods of 154,275,357 and 560 days on the 365-nm albedo.\\
\\
Long-term trends were also observed for the dayside winds of the upper clouds during the VEx mission. Between 2006--2013 the mean zonal velocity measured with ultraviolet images from the VMC camera displayed an increase from 80 to 100 m s$^{-1}$ at about 20$^{\circ}$S \citep[fig.~14 therein]{Khatuntsev2013}, while a similar increase trend was independently confirmed in this period with cloud tracking in VIRTIS-M images \citep{Hueso2015}. \citet{Kouyama2013} analyzed VEx/VMC images finding that the zonal winds equatorward of 30$^{\circ}$S exhibited a periodical perturbation of $\sim$10 m s$^{-1}$ every 257$\pm$2 terrestrial days which was ultimately interpreted as centrifugal waves \citep[fig.~4 therein]{Peralta2014b}. Measurements with Akatsuki/UVI images have also confirmed zonal windspeeds variations with time scales of $\sim$100 days between December 2015 and March 2017 \citep[figs.~8--10 therein]{Horinouchi2018}. Numerical simulations with GCMs have also been used to predict decadal variation in the zonal winds in Venus \citep{Parish2011}.\\
\\
The long-term variability of the winds at the level of the lower clouds was studied during the VEx mission on the nightside with VIRTIS-M by \citet{Hueso2012} and on the dayside with VMC by \citet{Khatuntsev2017}. During the years 2006--2008, the nightside winds at the deeper clouds showed at subpolar latitudes stronger variability than at lower ones \citep[figs.~7--8 therein]{Hueso2012}, while a comparison between the zonal winds during the Galileo flyby in 1990 and the 2-year average from VEx do not indicate noticeable variations except for the northern jet reported from the Galileo/NIMS images \citep[fig.~3 therein]{Sanchez-Lavega2017}. \citet{McGouldrick2017} reported an oscillation of approximately 150 days apparent between 30$^{\circ}$S--60$^{\circ}$S on the nightside clouds' radiance of the 1.74-$\mathrm{\mu m}$ VIRTIS-M images, with this time scale being consistent with the cycle of cloud formation and evolution driven by the radiative dynamical feedback and gravitational settling of clouds' particles. The analysis of the dayside winds with VMC from 2006 December to 2013 August ($\sim$1,200 days) revealed long-term variations on both components of the wind at $\sim$20$^{\circ}$S, with the meridional winds exhibiting a gradual increase until doubling its magnitude, while the zonal winds seemed subject to an apparent oscillation of $\sim$3 years \citep[fig.~11 therein]{Khatuntsev2017}.\\
\\
In order to perform an analysis of the variability of the zonal winds at the nightside lower clouds using an even longer time scale, we decided to combine data not only from Akatsuki/IR2 and VEx/VIRTIS, but also from ground-based observations and \textit{in situ} measurements. Figure \ref{figure:DecadalWinds} shows the variability of zonal winds at low latitudes (30$^{\circ}$S--30$^{\circ}$N) of the nightside lower clouds from 1978 December to 2017 February (spanning 38 years). The zonal winds during 1978 December (Fig.~\ref{figure:DecadalWinds}\textbf{A}) and 1985 June (Fig.~\ref{figure:DecadalWinds}\textbf{D}) correspond to the averages of instantaneous \textit{in situ} wind measurements between 50 and 60 km of altitude as provided by the Pioneer Venus Night probe \citep{Counselman1980} and VEGA landers \citep{Moroz1997}. To reduce the impact of transient phenomena like the equatorial jets, the zonal winds obtained with cloud tracking are presented as time averages of about 10--20 days in most of the cases, though this value strongly depends on the data availability and how this is distributed along time. Thus, zonal speeds from IRTF/SpeX images correspond to nearly instantaneous winds (several hours) while the average for 2004 May is provided by \citet{Limaye2006} for observations performed along 70 days. The mean zonal winds displayed in Fig.~\ref{figure:DecadalWinds}\textbf{A--I} were obtained from published results \citep{Counselman1980,Allen1984,Allen1987,Moroz1997,Crisp1989,Crisp1991b,Carlson1991,Chanover1998,Limaye2006}. The wind speeds from 2006 to 2017 correspond to a revisit of the VEx measurements (Fig.~\ref{figure:DecadalWinds}\textbf{J}) performed by \citet{Sanchez-Lavega2008,Hueso2012}, and our measurements with Akatsuki/IR2 (Fig.~\ref{figure:DecadalWinds}\textbf{M}), IRTF/SpeX (Fig.~\ref{figure:DecadalWinds}\textbf{L,N}) and TNG/NICS (Fig.~\ref{figure:DecadalWinds}\textbf{K}).\\

Although caution must be taken when comparing values from different time-averages and images acquired with different filters may provide some discrepancies in the vertical level sensed, Fig.~\ref{figure:DecadalWinds} suggests that the zonal wind speeds at the nightside lower clouds seems to experience long-term variation in the intensity of the winds of up to 30 m s$^{-1}$ (this total variation is larger than variabilities linked by the local time or longitudinal dependence described in subsections \ref{ssec:LTimeDep} and \ref{ssec:LongDep}). The zonal winds display a strong increase from the entry of the Pioneer Venus probes in 1978 (Fig.~\ref{figure:DecadalWinds}\textbf{A}) and the first cloud-tracked winds with ground-based observations (Fig.~\ref{figure:DecadalWinds}\textbf{B}) in the early 1980s. From 1983 to 1990 (Fig.~\ref{figure:DecadalWinds}\textbf{B}--\textbf{G}) the zonal winds exhibit a gradual decrease in magnitude, while from 2004 to 2008 the zonal winds experience an increase above the error bars (Fig.~\ref{figure:DecadalWinds}\textbf{J} from VEx data). This long-term behaviour of the nightside zonal winds from VIRTIS-M images diverges from the oscillation between 2007 and 2009 reported on the zonal winds between 15$^{\circ}$S--25$^{\circ}$S of the dayside middle-to-lower clouds found \citep[fig.~11a therein]{Khatuntsev2017}. The differences between the altitude range sensed in day and night side images \citep{Sanchez-Lavega2008,Takagi2011,Khatuntsev2017} or the distinct image sampling and time coverage for the winds obtained from VMC and VIRTIS-M may account for this discrepancy between VEx results.\\
\\
The decadal variability for the zonal winds at the lower clouds also keep similarities with the long-term trend for the SO$_2$ abundance at the cloud tops reported by \citep{Marcq2013NatGeo} during 30 years of observations combining Pioneer Venus and VEx observations. The gradual decrease of the zonal winds along 1983--1990 matches the decrease of the SO$_2$ at the upper clouds \citep[fig.~3 therein]{Marcq2013NatGeo}, what would support the idea that the vertical level sensed by the nightside images at 1.74, 2.26 and 2.32 $\mathrm{\mu m}$ might be variable. This would make sense if the photochemical activity at the upper clouds might could affect the total opacity of the cloud layer \citep{Knollenberg1980,Grinspoon1993PSS}. In such a case, faster zonal speeds may be coincident with periods when higher altitudes can be sensed in the nightside images.\\

\section{Conclusions}\label{sec:conclus}
We have presented  global results of the wind speeds at the nightside lower clouds of Venus during the first year of JAXA's Akatsuki mission. Both zonal and meridional winds were obtained applying cloud tracking techniques over  2.26-$\mathrm{\mu m}$ images acquired by the IR2 camera from a selection of 466 images spanning the time interval from 2016 March 22 to October 31. Automatic and manual measurements were applied independently to obtain 149,033 and 2,947 wind vectors, respectively. In the specific case of the manual measurements, the phase correlation technique was tested in the template matching for the first time on Venus with comparable results to those obtained using cross-correlation techniques in the space domain. No dependence is found between the wind speeds and the radiance or the size of the cloud tracers. This supports recent reports of wind variability due to strong horizontal shear and episodes of jets at the equator, and the identification of clouds' morphology consistent with shear instabilities. The meridional profiles of zonally-averaged wind speeds are in overall agreement with results during 2006--2008 from Venus Express but are systematically 10 m s$^{-1}$ faster. Our results confirm in the northern hemisphere the expected poleward decay of the zonal winds with symmetric winds between both hemispheres. Conversely to past observations with Venus Express, zonal speeds during the Akatsuki mission exhibits a local maximum in the westward windspeeds caused by either a local time dependence and/or influence of surface elevations, although the irregular coverage of the dataset prevents a definitive confirmation of its source. Also, a first analysis of the decadal wind variability is performed for zonal winds between 30$^{\circ}$S and 30$^{\circ}$N using a combination of \textit{in situ} and cloud tracking measurements performed between 1978 and 2017. Our results demonstrate yearly and decade wind variability, suggesting that the zonal winds at low latitudes might be affected by an oscillating disturbance with an amplitude of about $\sim$15 m s$^{-1}$ and a period of about 30 years.\\
\\
Finally, in order to facilitate future studies of wind variability and data comparison, we provide the full data set of wind measurements performed with the manual technique, accompanied by the template match for individual measurements and animations of the projected image sequences used towards this purpose (see accompanying Supplemental Material and Appendix \ref{sec:apdx}).\\

%% If you wish to include an acknowledgements section in your paper,
\section*{Acknowledgements}
J.P. acknowledges JAXA's International Top Young Fellowship. R.H. and A.S.-L. were supported by the Spanish MINECO project AYA2015-65041-P with FEDER, UE support and Grupos Gobierno Vasco IT-765-13. All authors acknowledge the work of the entire Akatsuki team. We are also grateful to the anonymous reviewer for his/her useful comments to improve the manuscript.\\% All the images used in this work can be downloaded from the public URLs included in the Supporting Information; any additional data may be obtained from the authors.

\appendix

\section{Description of the Supplemental Material}\label{sec:apdx}

The supplemental material that accompanies this work consists on a compressed file that contains not only the numerical values for our manual wind measurements obtained with the images of Akatsuki/IR2 camera, but also a large set of images exhibiting the morphology of cloud tracers and the quality of the template matching, geometrical projections of the IR2 images and animations of the cloud motions. The contents of the supplemental material are described as it follows:

\begin{itemize}
	\item a Readme file with a more detailed description of the contents of compressed file, as well as explanations about the naming of files and folders.
	\item a Microsoft Excel file with the numerical values of the 2,277 manual wind measurements and where the motions of the equatorial cloud discontinuity have been filtered out.
	\item a total of 103 folders for each pair of IR2 images, with wind measurements, animations and cloud tracers' morphology and position.
\end{itemize}

%\bibliography{bibliografia}



\end{document}